\documentclass{article}
\usepackage[utf8]{inputenc}
\usepackage{graphicx}
\usepackage[top=1in,left=1in,right=1in,bottom=1in]{geometry}
\usepackage{amsmath}
\usepackage{color}
\usepackage{dcolumn}
\usepackage{amsmath}
\usepackage{amsfonts}
\usepackage{amssymb}
\usepackage{bm}
\usepackage{epstopdf}
\usepackage{siunitx}
\usepackage{picture}
\usepackage{enumitem}
\usepackage{afterpage}
\usepackage{placeins}
\usepackage{float}
\usepackage{mathdesign}
\usepackage[english]{babel}
\usepackage{xr}
\graphicspath{ {./Figs/} }

\usepackage{url}
\usepackage[numbers,sort&compress]{natbib}

\providecommand{\bibcommenthead}{}

\makeatletter
\newcommand*{\addFileDependency}[1]{
  \typeout{(#1)}
  \@addtofilelist{#1}
  \IfFileExists{#1}{}{\typeout{No file #1.}}
}
\makeatother

\begin{document}
\title{Exceptional-point-like Sensing near Hermitian Critical Points}
\author{Jiang-Shan Tang$^{1,2,*}$, Long-Qi Xiao$^{1, }$\footnote{These authors contributed equally to this work.}\ , Hao-Dong Wu$^{1,3}$, Yuwei Jing$^{1}$, Han Zhang$^{4}$,\\ Ya-Ping Ruan$^{5}$, Wuming Liu$^{6,\dagger}$, Yan-Qing Lu$^{1,\dagger}$, Keyu Xia$^{1,7,8, }$\footnote{Email: wmliu@iphy.ac.cn, yqlu@nju.edu.cn, keyu.xia@nju.edu.cn}}

\date{}

\maketitle

\noindent $^1$College of Engineering and Applied Sciences, National Laboratory of Solid State Microstructures,Nanjing University, Nanjing, 210023, China .\\
$^2$College of Advanced Interdisciplinary Studies, National University of Defense Technology, Changsha, 410073, China.\\
$^3$Xi'an Institute of Applied Optics, Xi'an, 710065, China.\\
$^4$School of Physics, Nanjing University, Nanjing, 210023, China.\\
$^5$Laboratory of Solar and Space Instruments, Nanjing Institute of Astronomical Optics and Technology, Chinese Academy of Sciences, Nanjing, 210042, China.\\
$^6$Beijing National Laboratory for Condensed Matter Physics, Institute of Physics, Chinese Academy of Sciences, Beijing, 100190, China.\\
$^7$Hefei National Laboratory, Hefei, 230088, China.\\
$^8$Shishan Laboratory, Suzhou Campus of Nanjing University, Suzhou, 215000, China.\\

%\textbf{Abstract} 
{\bf A non-Hermitian system at an exceptional point (EP), a specific critical point (CP) associated with the parity-time symmetric phase transition, exhibits a sublinear response to perturbation and promise unprecedented sensitivity beyond the linear-response Hermitian sensors, so far operating at the diabolic points (DP). Despite great advancements, its sensitivity enhancement is fundamentally limited by the divergent Petermann factor, intrinsically rooted in the non-Hermitian eigenvector degeneracy, and practically by the system complexity. Here, we report the CP-resulting square-root response to the refractive index change and enhanced sensitivity in a simple chiral Hermitian cavity without phase transitions. Because of the inherent eigenvector orthogonality, this CP-based Hermitian sensor exhibits an EP-like response and enhanced sensitivity, breaking the Petermann-factor limit of sensitivity in non-Hermitian counterparts. This work paves the way towards exploring the Hermitian CPs for ultrasensitive sensing outperforming both the EP- and DP-based sensors.} \\

%\section*{Introduction}
%connection of phase transition and the cp
When a phase transition occurs in a physical system, some features display an abrupt change in parameter space, manifesting a critical point (CP). A system near a CP is very sensitive to a perturbation and can be utilized as an ultrasensitive sensor. By exploiting the parity-time ($\mathcal{PT}$) symmetric phase transition, a non-Hermitian sensor near the so-called exceptional point (EP), where its eigenvalues and eigenvectors simultaneously coalesce~\cite{CarlBanderPRP.70.947.2007, GeLiNPreview.11.2017,GanainyNatPhys.14.11.2018, Nat.Mat.18.783.2019}, has shown the essential advantages in response~\cite{JanPRL.112.203901.2014, FrancoPRL.117.110802.2016, AluScience.363.7709.2019, LanYangNature.548.192.2017, KerryVahalaNature.576.65.2019, MercedehNature.548.187.2017, MercedehNature.576.70.2019, DuScience.364.878.2019, KanteNPhysics2020,marenminNP.16.2020,AluReview.2023,GuoPRL.125.240506, LanYangSciAdv.10.eadl5037.2024, CWQiuPRL.130.227201, CWQiuNatElect.2.335,WangxuehuaPRL.2023,wangxuehuareview.2025} and sensitivity~\cite{XiaNatPhoton.19.109.2025, KenLiuNatNanoTech.19.1472.2024, KottosNature.607.697.2022} over a conventional Hermitian counterpart operating near the diabolic point (DP). Interplay of non-Hermitian physics and topology physics has shown EP-related topological structures~\cite{Nature.526.2015,Nature.537.2016,zhigangchensicence2021,kantescience.373.2021,JianZiPRL.135.046203, QihuangGongPRL.129.053903} and also promises extremely high sensitivity~\cite{Sci.Adv.10.eadp6905.2024}. Despite great advancements, the sensitivity of non-Hermitian sensors is strictly constrained practically by the complex experimental implementation and fundamentally by the divergent Petermann factor (PF), which intrinsically originates from the non-Hermitian degeneracy~\cite{VahalaNC.11.1610.2020, JanNC.11.2454.2020, nat.commum.9.2018}. The large PF factor can dramatically amplify some extra noises and severely limits the sensitivity enhancement. Here, we report a chiral Hermitian system for achieving the EP-like response and enhanced sensitivity in the absence of phase transitions, overcoming the experimental complexity and the PF problem of its non-Hermitian counterpart.

\begin{figure}%
	\centering
	\includegraphics[width=0.85\textwidth]{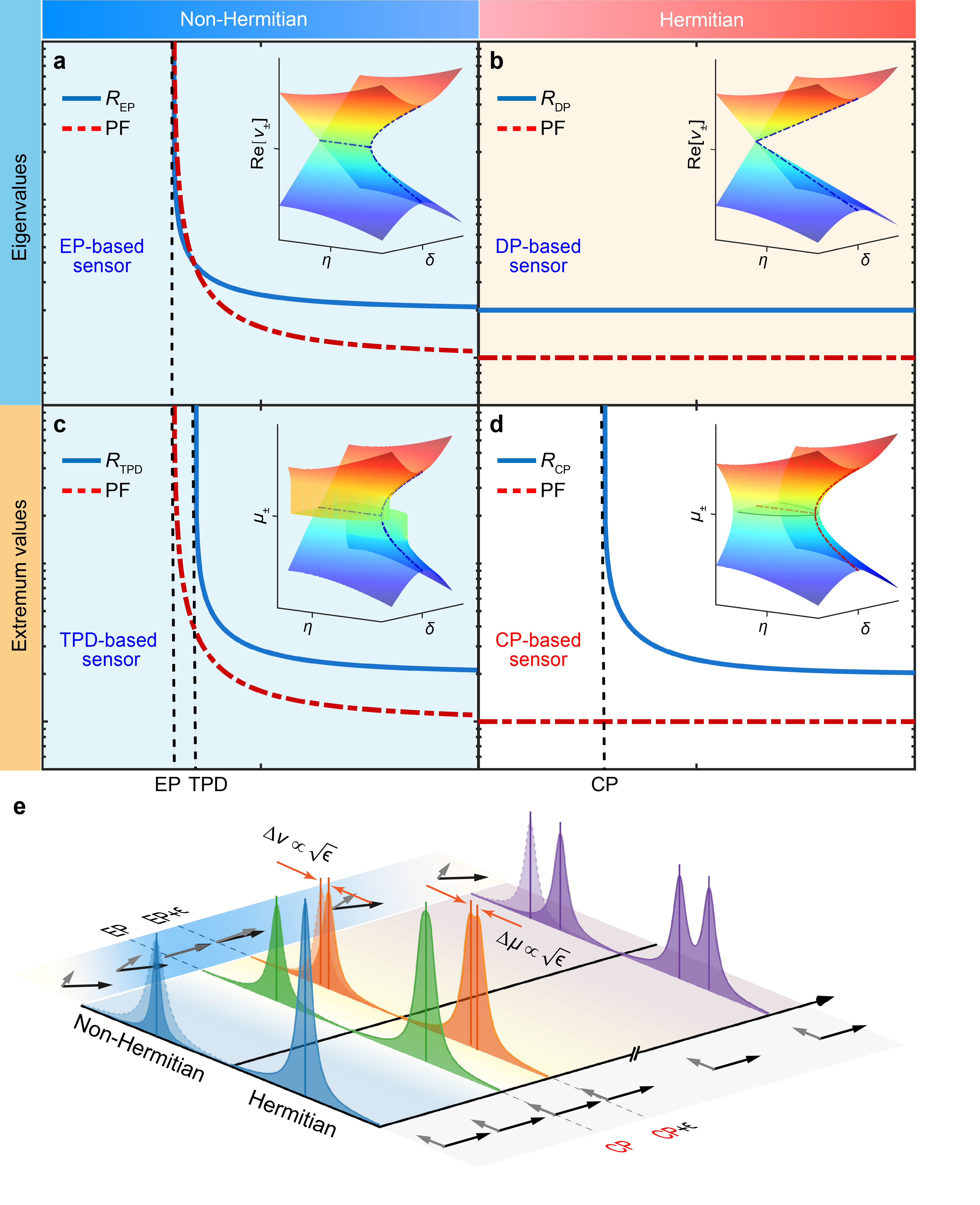}
	\caption{\textbf{Conceptual illustrations comparing sublinear response enhanced sensing in Hermitian and non-Hermitian systems.}
		\textbf{a}-\textbf{d}, Comparison of dynamic response $R$ (solid blue curves) and PF (dashed red curves) for the non-Hermitian EP-based sensor (\textbf{a}), the Hermitian DP-based sensor (\textbf{b}), the non-Hermitian TPD-based sensor (\textbf{c}) and our Hermitian CP-based sensor (\textbf{d}). The observable is the eigenfrequency splitting in the EP- and DP-based sensors, whereas it is the spectral-extremum splitting in the TPD- and CP-based sensors. Insets in \textbf{a} and \textbf{b} show the real parts of the system eigenvalues of the non-Hermitian and Hermitian sensors in the $\eta-\delta$ parameter space, respectively. Insets in  \textbf{c} and \textbf{d} depict the spectral extremum values with respect to their average value in the $\eta-\delta$ parameter space as \textbf{a} and \textbf{b}. The black vertical dashed lines indicate the position of the EP, TPD, and CP, respectively.
		\textbf{e,} Diagram comparing non-Hermitian EP-based and Hermitian CP-based sensors. Left, EP-based sensor with a panel illustrating nonorthogonal eigenvectors. The eigenmode transmission spectra are presented. Right, CP-based sensor with a panel illustrating orthogonal eigenvectors. The overall transmission spectra are presented.
	}\label{Fig:illustrations}
\end{figure}

% non-Hermitian sensor and EP
A conceptual comparison of the conventional Hermitian DP-based sensor, the non-Hermitian sensors and our CP-based sensor is illustrated in Fig.~\ref{Fig:illustrations}. These sensors involves two coupled modes. The figure depicts the measured frequency separation as a system response $\chi$ in the $\delta-\eta$ parameter space (inset surfaces), where $\delta$ is half of the frequency detuning between the two modes, the perturbation-induced coupling strength $\eta$ is proportional to the normalized perturbation $\epsilon$ ($\epsilon \ll 1$), the dynamic response $R$ (blue curves), defined as the derivative of the response with respect to perturbation, i.e. $R=\partial \chi/\partial\epsilon$. The PF (red dashed curves) is also presented for four sensors for comparison.
%\epsilon
A coupled two-mode system, as a general model of a sensor, can measure a perturbation at a bias $\epsilon_\text{b}$ as the induced frequency separation of observables.
% DP sensor
Conventionally, a Hermitian two-mode sensor with a vanishing gain/loss contrast, $\Delta\kappa = 0$, operates at the DP where $\delta = 0$ and $\epsilon_\text{b} = 0$. For any perturbation, its eigenfrequency splitting follows an $\epsilon$-dependence, $\Delta\nu = \text{Re}[\nu_+] - \text{Re}[\nu_-] \propto \epsilon$, as shown by the real parts $\text{Re}[\nu_\pm]$ of the two complex eigenfrequencies in Fig.~\ref{Fig:illustrations}\textbf{b}. This linear response implies a constant dynamic response, $R_\text{DP} = \partial\Delta\nu/\partial\epsilon = \text{const.}$, and sets a fundamental limitation to the available sensitivity in the DP regime of a Hermitian sensor.
% EP sensor
By exploring the abrupt response at the EP associated with the $\mathcal{PT}$-symmetric phase transitions, a non-Hermitian sensor can tackle this $\epsilon$-dependence limitation, see Fig.~\ref{Fig:illustrations}{\bf a}. Typically, it measures the eigenfrequency splitting, as shown in the left panel in Fig.~\ref{Fig:illustrations}{\bf e}.
Below the EP, the two eigenmode transmission spectra locate at the same frequency (Fig.~\ref{Fig:illustrations}\textbf{e}), implying a vanishing eigenfrequency splitting, see Figs.~\ref{Fig:illustrations}\textbf{a}.
Slightly beyond the EP, $\eta(\epsilon_\text{b}) \gtrsim \eta_\text{EP}$, the eigenfrequency splitting $\Delta \nu$ follows a $\sqrt{\epsilon}$-dependence for the on-resonance case $\delta =0$, see Fig.~\ref{Fig:illustrations}\textbf{e} and the inset in Fig.~\ref{Fig:illustrations}\textbf{a}. Therefore, the dynamic response (blue curve) is enhanced by a divergent factor $\propto 1/\sqrt{\epsilon}$ compared to a conventional Hermitian sensor, as shown in Fig.~\ref{Fig:illustrations}\textbf{a}. Well above the EP, the splitting is large, approaching a linear dependence on the perturbation.
Because of this remarkably enhanced dynamic response, non-Hermitian sensors have the potential to achieve unprecedented sensitivity overwhelming the conventional Hermitian counterpart~\cite{JanPRL.112.203901.2014, FrancoPRL.117.110802.2016}.
Despite of some debates~\cite{ShuChenPRL.131.160801, PhysRevLett.132.243601.2024}, this fundamental advantage of non-Hermitian sensors in sensitivity has been predicted by beautiful theories~\cite{JanPRL.112.203901.2014, FrancoPRL.117.110802.2016, PhysRevLett.123.180501.2019, OzdemirNC.13.3281, JianmingCaiPRL.124.020501}.

% PF problem in the gain-loss ep sensors
The sensitivity is more critical than the response for a sensor. It is crucially dependent on both the dynamic response and the noise in measurement~\cite{RMP.89.035002.2017}. The enhanced dynamic response have been successfully demonstrated in the gain-loss non-Hermitian sensors~\cite{LanYangNature.548.192.2017, KerryVahalaNature.576.65.2019, MercedehNature.548.187.2017, MercedehNature.576.70.2019}. In these active systems, one mode is amplified to balance the loss of the other. Inevitably, the strong gain adds the dominant noise to the system. Because the eigenvector coalescence, as indicated in Fig.~\ref{Fig:illustrations}\textbf{e},
noise in a non-Hermitian sensor can be equally amplified as its dynamic response by a divergent PF at the EP, see the red dashed curve in Fig.~\ref{Fig:illustrations}\textbf{a}. This large PF imposes constraint on real sensitivity enhancement and can compensate the sensitivity improvement in the worst case~\cite{VahalaNC.11.1610.2020, JanNC.11.2454.2020}, see the red dashed curve.
% TPD
This PF problem of a non-Hermitian sensor is partly tackled by measuring the frequency splitting near the transmission peak degeneracy (TPD),  where the spectrum starts to splits to a doublet-peak profile~\cite{KottosNature.607.697.2022}, see Fig.~\ref{Fig:illustrations}\textbf{c}.
The inset of Fig.~\ref{Fig:illustrations}\textbf{c} shows the peak positions $\mu_\pm$ of the transmittance spectra of a gain-loss non-Hermitian sensor.
Intriguingly, the $\sqrt{\epsilon}$-dependence response is obtain near the TPD. Because the TPD slightly shifts away from the EP~\cite{prj.9.1645.2021, KottosNature.607.697.2022}, the dynamic response is enhanced with a bigger factor than noise near the TPD. The PF-amplified noise is thus partly suppressed and a threefold enhancement is achieved for the signal-to-noise ratio.
% passive solutions but PF issue remains
A passive non-Hermitian sensor exhibits an important advantage over its active counterpart, because it completely avoids the strong gain-induced noise. It achieves the faithful enhancement in sensitivity~\cite{XiaNatPhoton.19.109.2025, KenLiuNatNanoTech.19.1472.2024}.  Nevertheless, the PF problem is yet to be overcome and thus severely limits the available sensitivity enhancement, because the intrinsic non-Hermitian degeneracy still remains. Despite highly demanded, it remains an outstanding challenge to develop a new sensing scheme to attain the EP-like enhancement in sensitivity without the PF limit.

% This work, CP-based Hamiltonian sensor
The concept of our CP-enhanced Hermitian sensor is depicted in Fig.~\ref{Fig:illustrations}\textbf{d} and is experimentally demonstrated in this work. The two modes equally decay with the same rate. Unlike the DP-based sensor observing the eigenfrequency splitting, our CP-based sensor measures the extremum separation of transmittance spectrum, $\Delta\mu$, see the right panel in Fig.~\ref{Fig:illustrations}\textbf{e}. Below a bias level that $\eta(\epsilon_\text{b}) \leq \eta_\text{CP}$, the transmission is a single peak. At $\eta_\text{CP}$, the transmission shows a plateau profile. Slightly biased above this point, the spectral-extremum splitting (SES) increases rapidly with the perturbation.
The inset in Fig.~\ref{Fig:illustrations}\textbf{d} shows the positions of two peaks of the transmission, $\mu_\pm$, versus the detuning and the perturbation. The two peaks locate at opposite sides of a spectral dip. The SES, defined as $\Delta\mu = \mu_+ - \mu_-$, displays a transition from a linear dependence to a square-root dependence on the perturbation at $\eta_\text{CP}$, clearly manifesting a CP. Comparing to the DP-based sensor, the dynamic response is enhanced by a factor of $\sqrt{\epsilon}$ near the CP. In stark contrast to the EP-based sensors, the eigenvectors of this CP-based Hermitian sensor are intrinsically orthogonal (see the right panel of Fig.~\ref{Fig:illustrations}\textbf{e}) and therefore the PF factor remains unity for varying perturbation (see the red dashed line in Fig.~\ref{Fig:illustrations}\textbf{d}), thoroughly overcoming the divergent PF problem.

\subsection*{Experimental setup}
The experimental setup of this CP-enhanced Hermitian sensor is illustrated in Fig.~\ref{Fig:setups}\textbf{a} (see Section 1 and Fig.~S1 for details of the setup in the Supplementary Material). Basically, the sensor is a chiral Fabry-P\'{e}rot cavity embedded with a magneto-optical (MO) crystal. The cavity is made from two mirrors with equal reflectivity of $99.5\%$. The effective length of the cavity is $L_c = 80~\rm{mm}$.
The MO medium is a ($111$)-cut cylindrical terbium gallium garnet (TGG) crystal, made by the company. The refractive index of the TGG crystal is $n_s \approx 1.95$, corresponding to a relative dielectric constant of $\varepsilon_s = n_s^2 \approx 3.8$. The length of the crystal is $L_\text{MO} = 20~\rm{mm}$ and its diameter is $3~\rm{mm}$, yielding a medium duty ratio of $\xi = L_\text{MO}/\left[ L_c + (\varepsilon_s -1) L_\text{MO} \right] \approx 14.7\% $. The anti-reflection dielectric film with equal transmittance of $99.7\%$ is coated on each end facet of the TGG crystal to reduce the loss of cavity mode. An electric coil surrounding the MO crystal is used to provide the external magnetic field $\mathbf{B}$. Due to the MO effect, $\mathbf{B}$ causes the different refractive index for the right circularly polarized (RCP) and the left circularly polarized (LCP) light fields. According to our measurement in~\cite{XiaNatPhoton.19.109.2025}, the Verdet constant of the TGG crystal is $C_\text{v} = 78.82~\rm{rad}/\rm{T}\cdot\rm{m}$.

A continuous-wave probe field with frequency $\nu$ from a laser (Toptica 795 \rm{nm}) is input into the cavity via the left mirror. It is linearly polarized and the polarization is controlled by a polarizer. The polarization is along the direction at an angle $\theta$ with the horizontal plane. We can tune the driving to the horizontally polarized (HP) and vertically polarized (VP) modes by adjusting the polarization orientation $\theta$.

Unlike the DP-based sensor detecting the eigenmode transmissions,  we measure the transmissions in the all-polarization (AP), HP and VP spaces as interference of the RCP and LCP components.
The transmitted light is first split into two paths by a beam splitter. A polarizing beam splitter is used again to divide the light in one path into the HP and VP modes. Light in the other path after the beam splitter is directly detected as the AP mode. The HP, VP and AP transmitted fields are measured by three photoelectric detectors. Note that all of three fields compose of the RCP and LCP components associated with different phases and amplitudes. By sweeping the frequency $\nu$ of the probe field around the central wavelength $\lambda \sim 795~\rm{nm}$, we can obtain the transmission spectra for measuring the perturbation.

As conceptually depicted in Fig.~\ref{Fig:illustrations}, our CP-based sensor overcomes the PF amplification of noise, compared to a non-Hermitian sensor.
To validate this concept, we examine the response and sensitivity enhancement by artificially adding intensity noise to the probe field and the output field.
We artificially add a strong intensity noise with a zero mean value to the probe field by randomly modulating the probe field via an acousto-optic modulator (AOM).
In practice, the photoelectric detectors also causes background noise due to the imperfect quantum efficiency and dark electric currents. Simulations of this type of noise is performed by adding a noise field randomly modulated in its amplitude into the output fields via a beam splitter~\cite{PengXuePRL.133.180801.2024} (see details in Supplementary Material).

\begin{figure}%
	\centering
	\includegraphics[width=1.0\textwidth]{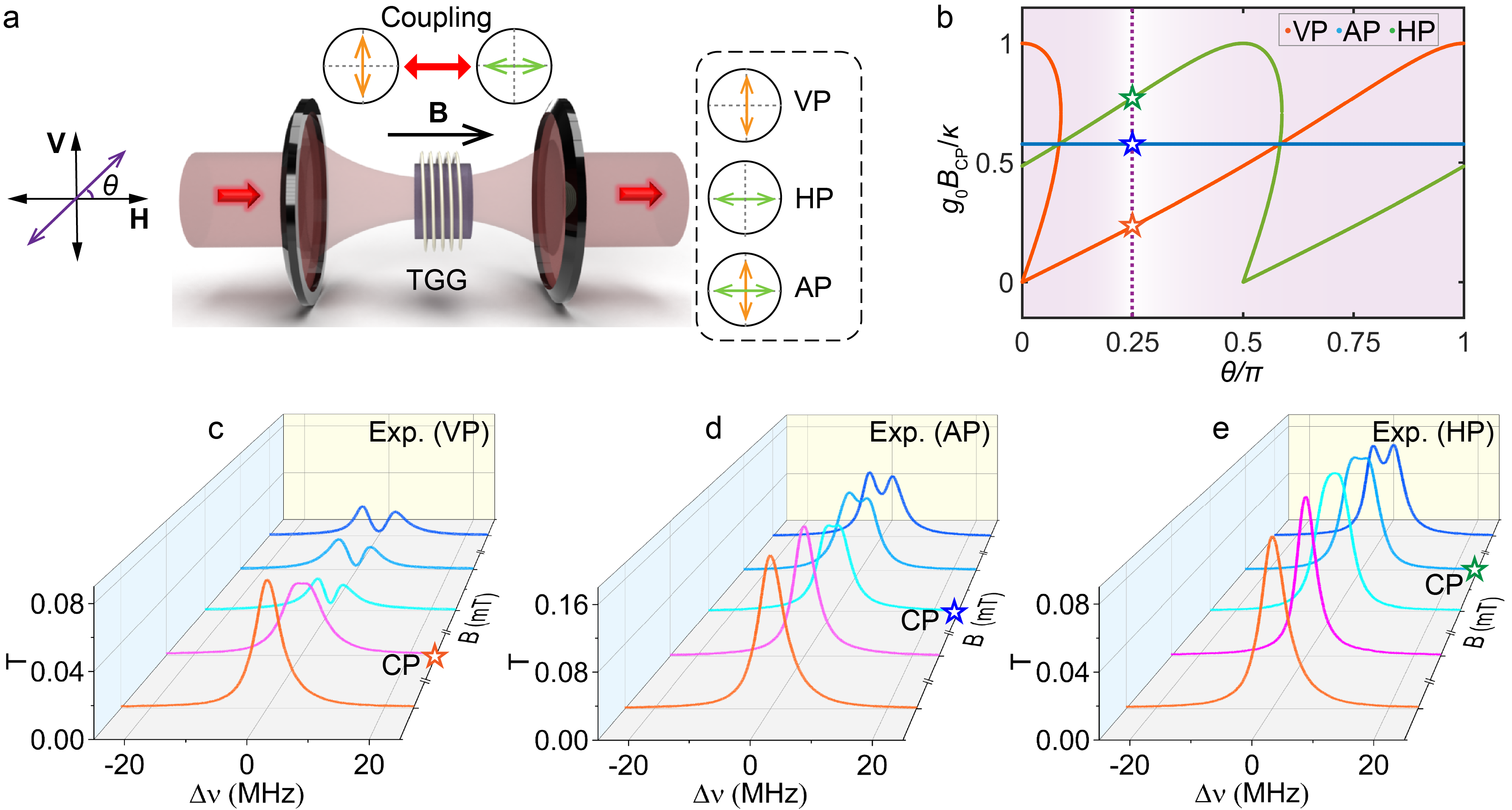}
	\caption{\textbf{Experimental setup and characterization.} \textbf{a} Schematic of the experimental setup for the CP-enhanced Hermitian sensor. TGG, terbium gallium garnet.
		\textbf{b,} CP position ($B_\text{CP}$) of the simulated VP (orange), AP (blue), and HP (green) transmission spectra as a function of the incident polarization angle $\theta$. The purple dashed line indicates the linearly polarized light incident at $\theta = \SI{45}{\degree}$, as marked by the orange, blue and green asterisks, respectively.
		\textbf{c}-\textbf{e}, Experimental transmission spectra for the VP (\textbf{c}), the AP (\textbf{d}), and the HP detection (\textbf{e}) under different magnetic fields.}\label{Fig:setups}
\end{figure}

\subsection*{Theory}\label{sec:Theory}
The Fabry-P\'{e}rot cavity supports two optical modes: the HP and VP modes, denoted as $\hat{a}_\text{H}$ and $\hat{a}_\text{V}$. We use $\nu_\text{H}$ and $\nu_\text{V}$ to denote their resonance frequencies, respectively.
The left and right mirrors causes external decay rates $\kappa_\text{H}^\text{ex}$ and $\kappa_\text{V}^\text{ex}$ to the HP and VP modes, respectively. Here, the external decays via the input and output mirrors are assumed to be equal for simplicity such that $\kappa_\text{H}^\text{ex} = \kappa_\text{V}^\text{ex} = \kappa^\text{ex}$. Considering the intrinsic losses $\kappa_\text{H}^\text{in}$ and $\kappa_\text{V}^\text{in}$, we respectively denote the total losses of the HP and VP modes as $\kappa_\text{H} = 2 \kappa^\text{ex} + \kappa_\text{H}^\text{in}$ and $\kappa_\text{V} = 2 \kappa^\text{ex} + \kappa_\text{V}^\text{in}$.
This HP and VP eigenmodes are linearly polarized and degenerate in resonance frequency such that $\nu_\text{H} = \nu_\text{V} = \nu_0$ and in dissipation such that $\kappa_\text{H} = \kappa_\text{V} = \kappa$,  leading to vanishing detuning $\delta = 0$ and gain/loss contrast $\Delta\kappa=0$.

A magnetic field $\mathbf{B}$ along the axis of TGG crystal can induce an interaction between the modes $\hat{a}_\text{H}$ and $\hat{a}_\text{V}$ with a strength $g = \varepsilon_s \xi C_\text{v} \nu_0/k_s$~\cite{XiaNatPhoton.19.109.2025}, where $k_s$ is the wavenumber of light in the MO medium that $k_s = 2\pi n_s/\lambda$, and $B$ is the amplitude of the magnetic field.
Then, the system is governed by the Hamiltonian:
\begin{equation}
	\label{eq:Hsys}
	H = \begin{bmatrix}
		\nu_\text{0} - i \kappa & i g B  \\
		- i g B & \nu_\text{0} - i \kappa
	\end{bmatrix}\;.
\end{equation}
Thus, this MO-induced chiral Fabry-P\'{e}rot cavity is a Hermitian two-mode sensor. It allows the DP- and CP-based measurement and thus facilitates comparison.

The MO-induced interaction converts the eigenmodes from linear polarization to circularly polarization such that $\hat{a}_+ = (\hat{a}_\text{H} + i \hat{a}_\text{V})/\sqrt{2}$ and $\hat{a}_- = (\hat{a}_\text{H} - i \hat{a}_\text{V})/\sqrt{2}$. The refractive index of the MO medium changes to $n_\text{R} = n_s \left(1+C_\text{v}B / 2 k_s\right)$ for the RCP eigenmode $\hat{a}_+$ and  $n_\text{L} = n_s \left( 1- C_\text{v}B / 2 k_s \right)$ for the LCP one $\hat{a}_-$, yielding a difference of $\Delta n = n_s C_\text{v}B / k_s \approx g B /  n_s \xi \nu_0$. This refractive-index difference causes the eigenfrequency splitting of the RCP and LCP eigenmodes, given by $\Delta = 2 g B = 2 n_s \xi \nu_0 \Delta n$.  In the RCP and LCP basis, $\hat{a}_+$ and $\hat{a}_-$, the Hermitian becomes $H^\prime = \left[\nu_0 + gB -i\kappa, 0; 0,  \nu_0 - gB -i\kappa \right]$.

The probe field drives the HP and VP modes with relative strengths $\cos\theta$ and $\sin\theta$.
% DP sensor
A DP-based sensor measures the splitting of two eigenmode transmissions as a DP-based sensor and yields a response $\nu = 2 gB$ and a dynamic response $R_\text{DP} = 2 g$.
% CP sensor
This work focus on the response and sensitivity enhancement of the CP-based sensor.
We can find the position of the peaks, namely the spectral extremum, by calculating the zeros of the derivative of the transmission $T$ with respect to the probe frequency.
When $\theta = \pi/4$, the SES in the VP, AP and HP spaces, as a response, are respectively given by
\begin{equation} \label{eq:Dmu}
	\renewcommand{\arraystretch}{2} % To adjust the line spacing
	\Delta \mu=\left\{\begin{array}{l}
		\text{Re}\left[2 \sqrt{-(\kappa-g B)^2+2 g B \sqrt{\kappa^2+(\kappa-g B)^2}}\right] \text{ in the VP space, } \\
		\text{Re}\left[2 \sqrt{-\left(\kappa^2+g^2 B^2\right)+2 g B \sqrt{\kappa^2+g^2 B^2}}\right] \quad \text{ in the AP space, } \\
		\text{Re}\left[2 \sqrt{-(\kappa+g B)^2+2 g B \sqrt{\kappa^2+(\kappa+g B)^2}}\right] \text{ in the HP space. }
	\end{array}\right.
\end{equation}
The two peaks locate at $\mu_\pm = \nu_0 \pm \Delta\mu/2$. %
The dynamic responses, defined as $R_\text{CP} = \partial \Delta\mu / \partial B$, show discontinuous transition in all three measurement scenarios. Below the CP, the response $\Delta\mu$ is null.
At $B_\text{CP}$, the splittings display an abrupt transition from null to a square-root dependence on the magnetic perturbation for $B > B_\text{CP}$ and thus the refractive-index difference $\Delta n$, as conceptually illustrated in Fig.~\ref{Fig:illustrations}\textbf{d}. Although phase transition is absent, this discontinuous derivative of the response with respect to the perturbation manifests a CP.
By setting $\Delta\mu = 0$, we find the  CP at $g B_\text{CP} \approx 0.23 \kappa$ in the VP measurement, $g B_\text{CP} = \sqrt{3}\kappa/3$ in the AP and $g B_\text{CP} \approx 0.77 \kappa$ in the HP.

Notably, the CP position exhibits a different dependence on the polarization orientation $\theta$. It is independent of $\theta$ in the AP measurement, while it shows complicate dependence on $\theta$ in the VP and HP cases, see Figs.~\ref{Fig:setups}\textbf{b}. We focus on the observation at $\theta = \pi/4$. The CP appears firstly in the VP measurement, then in the AP and finally in the HP. Below, we experimentally examine the CP-enhanced response and sensitivity at these three CPs (see Supplementary Material for details in experimental setups).

The more important sensitivity $S$ is defined as the ratio of measurement uncertainty or noise $\sigma$ to the dynamic response $R$ with in the total measurement $\tau$, namely $S = \sigma/R \sqrt{\tau}$~\cite{natphysics.21.8.2025}. In a gain-loss non-Hermitian sensor operating near the EP, the total noise basically includes five contributions arising from various sources. Its power can be written as:
\begin{equation}\label{eq:nonHermitiannoise}
	\sigma_{\text{EP}}^2 = \mathcal{PF}^2 \times \left(\sigma_{\text{gain}}^2 +\sigma_{\text{shot}}^2+ \sigma_\text{cl}^2\right) +R_{\text{EP}}^2\sigma_{\text{signal}}^2+ \sigma_{0}^2\;,
\end{equation}
where the gain-induced noise $\sigma_{\text{gain}}$ arises from spontaneous emission of the gain medium~\cite{siegman1986lasers, nat.commum.9.2018, PhysRevLett.123.180501.2019, WiersigPRJ2020}, and the shot noise $\sigma_\text{shot}$ originates from quantum vacuum fields and the photon projection noise of the probe field~\cite{PhysRevA.57.4736.1998}. These contributions represent the fundamental quantum noise. They are amplified by the EP-enhanced $\mathcal{PF}$, and $\mathcal{PF} \gg 1$ near the EP. In contrast, $\sigma_\text{cl}$ models the classical noise in the probe field such as the intensity fluctuation. This noise couples into the system through the input ports and is also likewise amplified by the PF~\cite{PhysRevA.57.4736.1998,nat.commum.9.2018}. In our experiment, we directly measure the output but not via the heterodyne detection. Both the shot noise and classical noise causes the similar uncertainty in measurement. Therefore, by adding artificial noise in the probe intensity, we can simulate the shot noise. The term  $\sigma_{\text{signal}}$ describes the noise  homologous to the signal. As a signal-induced output, it is related to the dynamic response. For an artificial signal, this noise can be the uncontrollable fluctuation or side-effect of the signal, e.g. such as heating noise. For a real signal from environment, this contribution disappears. The final term $\sigma_0$ reflects the background noise imposed by technical constraints including the imperfect quantum efficiency and dark electric currents of the photoelectric detectors. In an active non-Hermitian sensor, the gain-related noise is dominant and thus compensate the sensitivity enhancement. In contrast, the passive non-Hermitian sensor is advantaged because the noise $\sigma_\text{gain}$ is eliminated~\cite{XiaNatPhoton.19.109.2025}.

Remarkably, our Hermitian sensor contains no gain medium. Its eigenvectors are inherently orthogonal, completely avoiding the PF problem because $\mathcal{PF}$ remains unity. As a result, the total noise in a Hermitian sensor reduces to
\begin{equation}\label{eq:Hermitiannoise}
	\sigma_{\text{DP/CP}}^2 = \sigma_{\text{shot}}^2+ \sigma_{\text{cl}}^2+R_{\text{DP/CP}}^2\sigma_{\text{signal}}^2 + \sigma_{0}^2\;.
\end{equation}
In a non-Hermitian sensor, these two noise contributions are greatly amplified by the divergent PF near the EP. In our CP-based sensor, the noise amplification, defined as $G_\sigma =  \sigma_\text{CP}/\sigma_\text{DP}$, is eliminated  because the eigenmodes are orthogonal.
Clearly, when the signal-related contribution is negligible, leading to $\sigma_\text{CP} \approx \sigma_\text{DP}$, the sensitivity enhancement, evaluated as $G_S = S_\text{DP}/S_\text{CP} = G_\text{R} / G_\sigma$, approaches to the response improvement $G_R$. This is the case for measuring a real signal.
Notably, compared to a non-Hermitian sensor with a large PF and the strong gain-induced noise, our CP-based sensor can exhibit a much larger sensitivity enhancement even when two sensors display close dynamic response that $R_\text{EP} \sim R_\text{CP}$. This is a key advantage of our CP-based sensor over a non-Hermitian sensor.

\subsection*{Results}

\begin{figure}%
	\centering
	\includegraphics[width=0.9\textwidth]{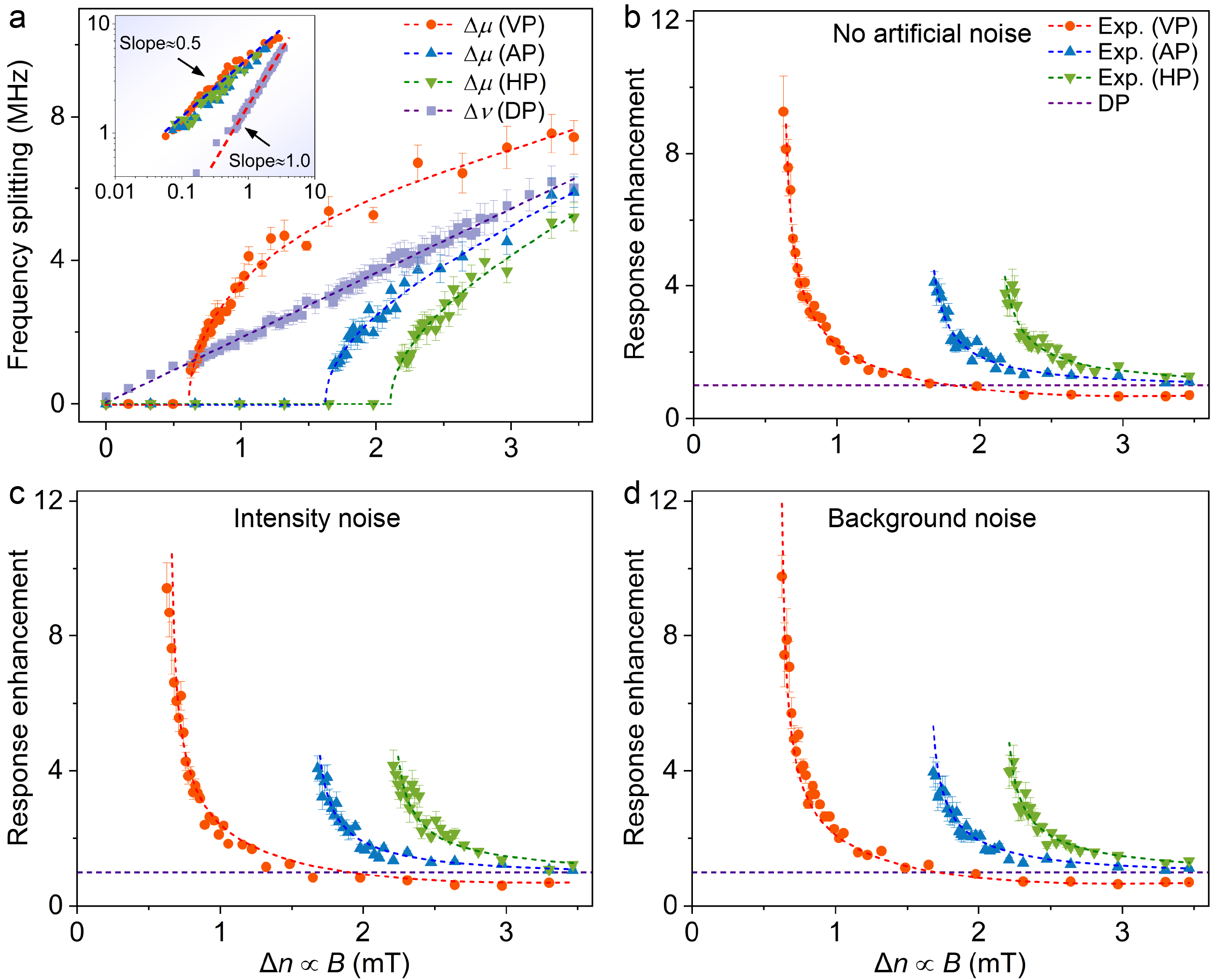}
	\caption{\textbf{Experimental observation and theoretical fitting of the CP-enhanced response to a refraction perturbation.} \textbf{a,} Eigenfrequency and spectral-extremum splittings versus the magnetic field. The insert shows the same data in a double-logarithmic fashion. \textbf{b}-\textbf{d}, Response enhancement, $G_\text{R}$, of the CP sensor relative to the DP sensor without additional noise (\textbf{b}), with added laser intensity noise (\textbf{c}), and with background noise (\textbf{d}). The discrete data points and curves represent the experimental and fitting results, respectively. }\label{Fig:response}
\end{figure}

% transmission spectra
The experimental transmission spectra for $\theta = \pi/4$ shown in Figs.~\ref{Fig:setups}\textbf{c}-\textbf{e} confirm the depicted concept for the CP-based sensor in Fig.~\ref{Fig:illustrations}\textbf{e}.
For $\mathbf{B} < B_\text{CP}$, corresponding to a small $\Delta n$, the VP, AP and HP transmissions exhibit singlet-peak structures, implying a null SES. This is because the eigenfrequency splitting is smaller than the eigenmode linewidths. At the CP of $B_\text{CP}$, the transmissions exhibit a plateau structure. As $B$ further increases, $\Delta n$ becomes bigger, causing a doublet-peak transmission. The spectral dip always appears at $\nu_0$. The CP is first observed in the VP spectrum, secondly in the AP and then in the HP, in good agreement with the theory (see the asterisks in Fig.~\ref{Fig:setups}\textbf{b}).

% response Figure 3
The responses of the DP-based and CP-based sensor are shown in Fig.~\ref{Fig:response}\textbf{a}. The DP-based sensor displays a linear response (purple), well fitted by a line. In contrast, all three responses in the VP, AP and HP space exhibit an abrupt change from null to a square-root dependence on the magnetic field $B$, see the log-log plot in the inset. Above $B_\text{CP}$, the experimental responses are well fitted by the theoretical results given in Eq.~\ref{eq:Dmu}. The fitting parameter $g$ takes the values of $0.90$, $0.89$, $0.89$ and $0.90~\rm{MHz}/\rm{mT}$ for the DP sensor, the VP, AP and HP measurements of the CP sensor, respectively. These close values indicate a good agreement between the experimental observations and theoretical predictions. The average value is $\bar{g} = 0.90~\rm{MHz}/\rm{mT}$. It means that a  magnetic field of $1~\rm{mT}$ can result in a refractive index difference of $\Delta n \approx 8.3\times 10^{-9}$.  Note that the SES in the VP measurement can surpass the eigenfrequency splitting in the DP-based sensor. This feature is not observed in the EP- and TPD-based sensors.

In comparison with the DP-sensor, the response enhancement is clearly observed in the CP-based measurements and is evaluated as $G_\text{R} = R_\text{CP}/ R_\text{DP}$, see Figs.~\ref{Fig:response}\textbf{b}-\textbf{d}.
Figure~\ref{Fig:response}\textbf{b} shows the enhancement factors without artificially adding noise, reaching maximal values of $G_\text{R} \approx 9.26$, $4.10$, and $4.02$ in the VP, AP and HP measurements, respectively.
As shown in  Fig.~\ref{Fig:response}\textbf{c} for artificially adding the intensity noise, the maximum enhancement factors in the VP, AP and HP spaces are $9.42$, $4.07$, and $4.15$, respectively. In simulation of strong background noise, the responses are enhanced by factors of $9.76$ in the VP measurement, $3.96$ in the AP and $4.27$ in the HP, respectively.
The experimental results are well fitted by the same function derived from Eq.~\ref{eq:Dmu} in each measurement scenario (see Methods). As indicated by the theoretical model, the experimentally observed responses are similar in all three cases, meaning that they are independent on the noise.

\begin{figure}[h]%
	\centering
	\includegraphics[width=0.9\textwidth]{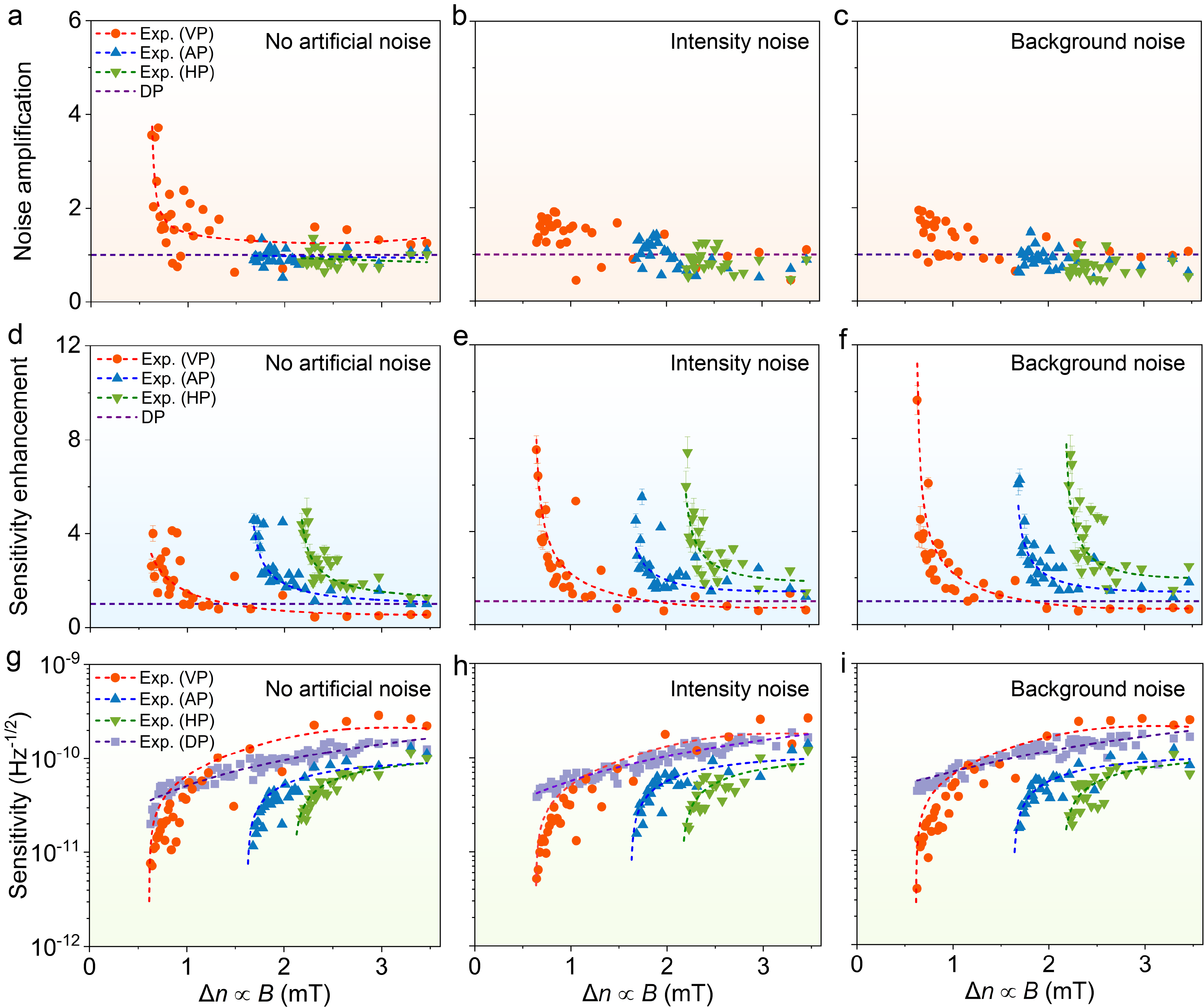}
	\caption{\textbf{Comparison of noise and sensitivity.} Noise amplification (\textbf{a}-\textbf{c}), sensitivity enhancement (\textbf{d}-\textbf{f}) and sensitivity (\textbf{g}-\textbf{i}) of the CP-based sensors compared to the conventional DP-based sensor in the absence of artificial noise (left panel), with added laser intensity noise (middle panel), and with background noise (right panel). Red, blue, and green data points correspond to the VP, AP, and HP measurements, respectively. Experimental data for the sensitivity, sensitivity enhancement and the noise amplification without artificial noise are fitted (see Methods for details).}\label{Fig:sensitivity}
\end{figure}

% noise and sensitivity
Figure~\ref{Fig:sensitivity} compares the noise,  and the sensitivity of the CP- and DP-based sensors in three cases of no artificial noise (left panel), adding intensity noise (middle panel) and background noise (right panel). The lines and data for the DP-based sensor are presented for comparison.
In the absence of artificial noise, the overall noise is amplified by a factor up to $G_\sigma \sim 3.7$ in the VP measurement, primarily attributed to the signal-related noise, $R_\text{CP}\sigma_{\text{signal}}$. Correspondingly, the sensitivity enhancement $G_S$ is about $4.10$ (Fig.~\ref{Fig:sensitivity}\textbf{d}). In our experiment, the probe laser field is input to the sensor via an AOM, for the purpose of reasonable comparison. The current noise in the AOM also causes fluctuation to the probe laser intensity. This noise is dominant over the contribution of $R_\text{CP}\sigma_{\text{signal}}$ in the AP and HP measurements. As a result, the noise is slightly enhanced by $\sim 1.2$ times. Although the response enhancement is smaller in these two cases of the AP and HP measurements, the noise level is lower than the VP measurement. Thus, the sensitivity is improved near the CP by a factor of $G_\text{S} \sim 4.41$, very close to the factor in the VP measurement. Without placing the AOM after the probe laser, the influence of the signal-related noise contribution becomes stronger. The noise amplification increases to about $1.60$ near the CP in the AP and HP measurements, although the values still smaller than the VP measurement (see Supplementary Material).

We experimentally examine the prediction of the PF factor by adding intensity noise with zero mean value to the probe field via an AOM driven by a uniformly distributed random current (see Supplementary Material for details). In our experiment, both the DP- and CP-based sensors directly measures the intensity of transmitted fields. Thus, the intensity noise of the probe laser field has the same effect as the shot noise and thus can also simulate the shot noise. As shown in Fig.~\ref{Fig:sensitivity}\textbf{b}, when the intensity noise dominates other contributions, the noise amplification factors are close to unity in all three measurement scenarios, clearly confirming that $\mathcal{PF} =1$. The sensitivity enhancement reaches larger values of $7.51$, $5.49$ and $7.40$ in the VP, AP and HP measurements, respectively.

% adding background noise
In practice, the photoelectric detectors also add background noise due to the imperfect quantum efficiency and dark electric currents. To simulation of the background noise, we add a noisy laser field randomly modulated in amplitude to the output field via a beam splitter as in~\cite{PengXuePRL.133.180801.2024} (see details in Supplementary Material). In this simulations, the artificial background noise is dominant. The noise amplifications display almost the same level as in the case of adding intensity noise (Fig.~\ref{Fig:sensitivity}\textbf{c}). The sensitivity enhancements are close to the response improvement factors (Fig.~\ref{Fig:sensitivity}\textbf{f}), reaching a maximum of about $9.64$ in the VP measurement.

%sensitivity
Figures~\ref{Fig:sensitivity}\textbf{g}-\textbf{i} show the sensitivities. It is clear that the CP-based sensor exhibits better sensitivities in all measurement scenarios than the DP-based sensor. Particularly, the VP measurement achieves a refraction sensitivity dramatically superior to $10^{-11}/\sqrt{\rm{Hz}}$.

%\FloatBarrier

\subsection*{Conclusions and outlook}\label{sec:conclusions}
We have demonstrated CP-enhanced sensing of refraction in a chiral and passive Hermitian optical system. This CP-based sensor exhibits an EP-like response and sensitivity well beyond the DP-based counterpart. Remarkably, our sensor tackles the gain-induced noise and the PF problem strongly constraining a non-Hermitian sensor. Such a general sensing method can be applied to diverse bipartite-coupled platforms and is expected to significantly advance the ongoing sensor revolution~\cite{sensor.revolution.2004}. Our sensor is very simple but ultrasensitive to a small change of the refractive index. The sensitivity can be further improved by three to four orders if the laser central frequency is fast swept with an AOM.
In combination with extra stabilization reducing technical noises, this work paves the way for facilitating the test of the standard model and searching for axions~\cite{PhysRevA.62.013815.2000,RevModPhys.82.557.2010}, which cause a very tiny change to the refraction index of vacuum.

%\begin{appendices}
\section*{Methods}\label{sec:method}
\noindent\textbf {Hamiltonian and transmission spectrum}\\
The system devices and parameters have been introduced in Sec.~\ref{sec:Theory}. A linearly polarized laser field with frequency $\nu$ excites the MO Fabry-P\'{e}rot cavity with a drive amplitude $\alpha$. Its polarization orientates at an angle $\theta$ to the horizontal plane. To facilitate the derivation of the transmission spectrum, we rewrite the system Hamiltonian in a quantum framework while explicitly including the external driving. In the frame rotating at frequency $\nu$,  the system Hamiltonian reads (we set the reduced Planck constant $\hbar=1$)
\begin{equation}\label{seq:Hamiltonian}
	\begin{split}
		\hat{H} = & -\left(\Delta-\delta\right) \hat{a}^\dagger_{\text{H}} \hat{a}_{\text{H}} - \left(\Delta+\delta\right) \hat{a}^\dagger_{\text{V}} \hat{a}_{\text{V}} + i gB\left(\hat{a}^\dagger_{\text{H}}\hat{a}_{\text{V}}-\hat{a}^\dagger_{\text{V}}\hat{a}_{\text{H}}\right) \; \\
		& +i\sqrt{2\kappa^{\text{ex}}}\alpha \cos\theta\left(\hat{a}^\dagger_{\text{H}}-\hat{a}_{\text{H}}\right) + i\sqrt{2\kappa^{\text{ex}}}\alpha \sin\theta\left(\hat{a}^\dagger_{\text{V}}-\hat{a}_{\text{V}}\right) \;,
	\end{split}
\end{equation}
where we have defined $\nu_0=\left(\nu_{\text{H}}+\nu_{\text{V}}\right)/2$, $\delta=\left(\nu_{\text{H}}- \nu_{\text{V}}\right)/2$, and $\Delta=\nu-\nu_0$. In Eq.~\eqref{seq:Hamiltonian}, the first two terms correspond to the free Hamiltonian, the third term describes the photon interaction, and the last two terms account for the external driving.

The Langevin equations of motion are given by
\begin{subequations}
	\label{seq:motionequation}
	\begin{align}
		\frac{d \hat{a}_\text{H}}{d t} &= i\left[\hat{H}, \hat{a}_{\text{H}} \right]+\mathcal{L}\left[\hat{a}_{\text{H}}\right]=\left(i\Delta_{\text{H}}-\kappa_{\text{H}}\right)\hat{a}_{\text{H}} + gB \hat{a}_{\text{V}}+\sqrt{2\kappa^{\text{ex}}}\alpha \cos\theta \; , \\
		\frac{d \hat{a}_\text{V}}{d t} &= i\left[\hat{H}, \hat{a}_{\text{V}} \right]+\mathcal{L}\left[\hat{a}_{\text{V}}\right]=\left(i\Delta_{\text{V}}-\kappa_{\text{V}}\right)\hat{a}_{\text{V}} - gB \hat{a}_{\text{H}}+\sqrt{2\kappa^{\text{ex}}}\alpha \sin\theta \; .
	\end{align}
\end{subequations}
Here,
$\mathcal{L}[\hat{o}] = \kappa(2\hat{o}^\dagger \hat{Q} \hat{o} -\hat{o}^{\dagger}\hat{o}\hat{Q} - \hat{Q} \hat{o}^{\dagger}\hat{o})$ with $\hat{o} = \{\hat{a}_\text{H}, \hat{a}_\text{V} \}$ is the Lindblad superoperator acting on an arbitrary system operator $\hat{Q}$. Setting Eq.~\eqref{seq:motionequation} to zero yields the system's steady-state solution:
\begin{subequations}
	\label{seq:steadystatesolution}
	\begin{align}
		\langle\hat{a}_\text{H}\rangle &= - \frac{\sqrt{2\kappa^{\text{ex}}}\alpha\left[\left(i\Delta+i\delta-\kappa_{\text{V}}\right)\cos\theta-gB\sin\theta\right]}
		{\left(i\Delta-i\delta-\kappa_{\text{H}}\right)\left(i\Delta+i\delta-\kappa_{\text{V}}\right)+\left(gB\right)^2} \; , \\
		\langle\hat{a}_\text{V}\rangle &= - \frac{\sqrt{2\kappa^{\text{ex}}}\alpha\left[gB\cos\theta+\left(i\Delta-i\delta-\kappa_{\text{H}}\right)\sin\theta\right]}
		{\left(i\Delta-i\delta-\kappa_{\text{H}}\right)\left(i\Delta+i\delta-\kappa_{\text{V}}\right)+\left(gB\right)^2} \; .
	\end{align}
\end{subequations}
The output fields follow from the input-output relations, $\hat{a}_{\text{H,out}}=\sqrt{2\kappa^{\text{ex}}}\hat{a}_{\text{H}}$ and $\hat{a}_{\text{V,out}}=\sqrt{2\kappa^{\text{ex}}}\hat{a}_{\text{V}}$. The VP, AP, and HP transmissions are defined as $T_{\text{VP}}=\lvert \langle\hat{a}_\text{V,out}\rangle\rvert^2/\lvert \alpha\rvert^2$, $T_{\text{AP}}=\lvert \langle\hat{a}_\text{H}\rangle\mathbf{e_{\text{H}}}+\langle\hat{a}_\text{V}\rangle\mathbf{e_{\text{V}}}\rvert^2/\lvert \alpha\rvert^2$, and $T_{\text{HP}}=\lvert \langle\hat{a}_\text{H}\rangle\rvert^2/\lvert \alpha\rvert^2$, where $\left\{\mathbf{e_{\text{H}}},\mathbf{e_\text{V}}\right\}$ represent the HP and VP unitary vectors satisfying $\mathbf{e_i}\cdot\mathbf{e_j}=\delta_{i,j}$ with $i,j=\text{H,V}$.

We first consider the Hermitian case, where $\kappa_{\text{H}} = \kappa_{\text{V}} = \kappa$. The eigenvalue of the system can be easily obtained as
\begin{equation}\label{seq:Hermitianeigenvalue}
	\nu_{\pm} = \nu_0 - i\kappa \pm \sqrt{g^2B^2 + \delta^2} \;.
\end{equation}
From Eq.~\eqref{seq:Hermitianeigenvalue}, one can plot a three-dimensional eigenvalue surface in the parameter space spanned by $\eta$ and $\delta $ (Fig.~\ref{Fig:illustrations}\textbf{b}), where $\eta=gB$. For $\delta = 0$, the eigenfrequency splitting exhibits a linear dependence on the coupling strength $gB$.

The transmission of the CP-based sensor is given by
\renewcommand{\arraystretch}{3} % To adjust the line spacing
\begin{equation} \label{seq:transmission}
	T (\Delta) =\left\{\begin{array}{l}
		\dfrac{\left(2\kappa^{\text{ex}}\right)^2\left[\left(gB\cos\theta-\kappa\sin\theta\right)^2+ \left(\Delta-\delta\right)^2\sin^2\theta\right]}
		{\left(g^2B^2+\kappa^2-\Delta^2+\delta^2\right)^2+4\kappa^2\Delta^2} \text{ in the VP space,} \\
		\dfrac{\left(2\kappa^{\text{ex}}\right)^2\left(g^2B^2+\kappa^2+\Delta^2+\delta^2+2\Delta\delta\cos 2\theta\right)}
		{\left(g^2B^2+\kappa^2-\Delta^2+\delta^2\right)^2+4\kappa^2\Delta^2} \quad \text{ in the AP space,} \\
		\dfrac{\left(2\kappa^{\text{ex}}\right)^2\left[\left(\kappa\cos\theta+gB\sin\theta\right)^2+\left(\Delta+\delta\right)^2 \cos^2\theta\right]}
		{\left(g^2B^2+\kappa^2-\Delta^2+\delta^2\right)^2+4\kappa^2\Delta^2} \text{ in the HP space.}
	\end{array}\right.
\end{equation}
By solving $\partial T/\partial \nu=0$, one can construct the three-dimensional extremal surface of the transmission spectrum in the $\eta-\delta$ parameter space. The inset in Fig.~\ref{Fig:illustrations}\textbf{d} shows the extremal surface of the transmission spectrum in the AP space. It is worth noting that, when $\delta = 0$, there always exists a common extremum $\mu_{0, \text{m}} = \nu_0$ with $(\text{m} = \text{VP} , \text{AP}, \text{HP})$ independent of the parameter $gB$. However, as $gB$ increases, two additional maxima emerge at $\mu_{+, \text{m}}$ and $\mu_{-,\text{m}}$. They symmetrically appear on both sides of $\mu_{0,\text{m}}$ in the transmission spectrum, forming a double-peak profile. The number of extrema abruptly jumps from one to three, manifesting a system CP. The position of the CP differs in the VP, AP, and HP measurements.

In the VP measurement scenario, the maxima are
\begin{equation}\label{seq:extremaV2}
	\mu_{\pm,\text{VP}}=
	\begin{cases}
		\nu_0 \pm \sqrt{g^2 B^2-\kappa^2}, & \sin\theta=0,\\
		\nu_0 \pm \left[\frac{-(\kappa \sin\theta-g B \cos\theta)^2 + g B \sqrt{\mathcal{N}_{\text{VP}}(\theta)}}{\sin^2 \theta}\right]^{1/2}, & \sin\theta\neq 0,
	\end{cases}
\end{equation}
where $\mathcal{N}_{\text{VP}}(\theta)=g^2B^2+4\kappa^2\sin^2\theta-4\kappa gB \sin\theta \cos\theta$. The extremum is a real number. When $\sin\theta=0$, the CP occurs at $gB_{\text{CP}} = \kappa$, where the number of extrema sudently changes from one to three. When $\sin\theta\neq 0$, the CP position is determined by numerically solving $\mu_{+,\text{VP}}-\nu_0=\mu_{-,\text{VP}}-\nu_0=0$, as shown in the orange solid curve in Fig.~\ref{Fig:setups}\textbf{b}.

In the HP measurement scenario, the maxima are
\begin{equation}\label{seq:extremaH2}
	\mu_{\pm,\text{HP}} =
	\begin{cases}
		\nu_0 \pm \sqrt{g^2B^2 - \kappa^2}, & \cos\theta = 0, \\[8pt]
		\nu_0 \pm \left[\frac{
			-(\kappa \cos\theta + gB\sin\theta)^2 + gB\sqrt{\mathcal{N}_{\text{HP}}(\theta)}
		}{\cos^2\theta}\right]^{1/2}, & \cos\theta \neq 0,
	\end{cases}
\end{equation}
where $\mathcal{N}_{\text{HP}}(\theta)
= g^2B^2 + 4\kappa^2\cos^2\theta + 4\kappa gB\sin\theta\cos\theta$. Similarly, for $\cos\theta = 0$, the CP occurs at $gB_{\text{CP}} = \kappa$, while for $\cos\theta \neq 0$, its position is obtained by numerically solving $\mu_{+,\text{HP}}-\nu_0=\mu_{-,\text{HP}}-\nu_0=0$, as indicated by the green solid curve in Fig.~\ref{Fig:setups}\textbf{b}.

In the AP measurement scenario, the maxima are
\begin{equation}\label{seq:extremaA1}
	\mu_{\pm,\text{AP}} = \nu_0 \pm \sqrt{-\left(\kappa^2+g^2B^2\right)+2 g B \sqrt{g^2 B^2+\kappa^2}} \;.
\end{equation}
The CP always occurs at $g B_{\text{CP}}=\kappa/\sqrt{3}$, independent of $\theta$ (see the blue solid line in Fig.~\ref{Fig:setups}\textbf{b}).

The SES in the CP-based sensor is defined as $\Delta\mu_{\text{m}} = \mu_{+, \text{m}}-\mu_{-, \text{m}} (\text{m} = \text{VP}, \text{AP}, \text{HP})$.  It follows that the SESs in all three measurement scenarios scale as the square root of a perturbation $\epsilon$ near the critical point $gB_{\text{CP}}$. In this work, we leverage this sublinear response for CP-based sensing with enhanced response and sensitivity performance.

Next, we turn to the non-Hermitian case, where $\kappa_{\text{H}} \neq \kappa_{\text{V}}$, which allows for the emergence of EPs and markedly different spectral behavior compared with the Hermitian case. In our MO Fabry-P\'{e}rot cavity, this can be constructed either by introducing an additional loss contrast between the two cavity modes (e.g., using a liquid crystal~\cite{XiaNatPhoton.19.109.2025}) or by adding gain for one of the modes. The eigenvalue of the non-Hermitian system is given by
\begin{equation}\label{seq:nonHermitianeigenvalue}
	\nu_{\pm} = \nu_0 - i\frac{\kappa_{\text{H}} + \kappa_{\text{V}}}{2} \pm \sqrt{g^2B^2 + \left(\delta-i\frac{\kappa_{\text{H}} - \kappa_{\text{V}}}{2}\right)^2} \;.
\end{equation}
Equation~\eqref{seq:nonHermitianeigenvalue} yields a three-dimensional eigenvalue surface in the $\eta-\delta$ parameter space, as shown in Fig.~\ref{Fig:illustrations}\textbf{a}. For $\delta = 0$, the EP occurs at $gB_{\text{EP}}=(\kappa_{\text{H}} - \kappa_{\text{V}})/2$. In contrast to Hermitian systems, near the EP, i.e., $gB = gB_{\text{EP}} + \epsilon$, the eigenfrequency splitting exhibits a $\sqrt{\epsilon}$ dependence. This feature motivates the design of the EP-based sensing scheme~\cite{JanPRL.112.203901.2014, FrancoPRL.117.110802.2016}.

In the non-Hermitian configuration, the transmission in the TPD-based sensor can be expressed as
\renewcommand{\arraystretch}{3} % To adjust the line spacing
\begin{equation} \label{seq:nontransmission}
	T (\Delta) =\left\{\begin{array}{l}
		\dfrac{\left(2\kappa^{\text{ex}}\right)^2\left[\left(gB\cos\theta-\kappa_{\text{H}}\sin\theta\right)^2+ \left(\Delta-\delta\right)^2\sin^2\theta\right]}
		{\mathcal{D}(\Delta)} \text{ in the VP space,} \\
		\dfrac{\left(2\kappa^{\text{ex}}\right)^2\left[\mathcal{N}_{\text{AP}}\left(\theta\right)+\Delta^2+\delta^2+2\Delta\delta\cos 2\theta\right]}
		{\mathcal{D}(\Delta)} \quad \text{ in the AP space,} \\
		\dfrac{\left(2\kappa^{\text{ex}}\right)^2\left[\left(\kappa_{\text{V}}\cos\theta+gB\sin\theta\right)^2+\left(\Delta+\delta\right)^2 \cos^2\theta\right]}
		{\mathcal{D}(\Delta)} \text{ in the HP space,}
	\end{array}\right.
\end{equation}
where $\mathcal{N}_{\text{AP}}\left(\theta\right)=g^2B^2 + \kappa_{\text{V}}^{2}\cos^2\theta+\kappa_{\text{H}}^{2}\sin^2\theta+gB\left(\kappa_{\text{V}}-\kappa_{\text{H}}\right)\sin 2\theta$ and $\mathcal{D}(\Delta) = \left(g^2B^2+\kappa_{\text{H}}\kappa_{\text{V}}-\Delta^2+\delta^2\right)^2+\left[\Delta\left(\kappa_{\text{H}}
+\kappa_{\text{V}}\right)-\delta\left(\kappa_{\text{H}}-\kappa_{\text{V}}\right)\right]^2$. Evidently, for $\kappa_{\text{H}} = \kappa_{\text{V}} = \kappa$, Eq.~\eqref{seq:nontransmission} simplifies to Eq.~\eqref{seq:transmission}. By solving $\partial T/\partial \nu = 0$, the extremal surface of the transmission spectrum in the AP space can be determined in the $\eta-\delta$ parameter space (Fig.~\ref{Fig:illustrations}\textbf{c}). For $\delta = 0$, the transmission spectrum shows a TPD~\cite{prj.9.1645.2021, KottosNature.607.697.2022}. Near this TPD, the extremum splitting also exhibits a square-root dependence on the perturbation $\epsilon$. Although the TPD-based sensor offers higher sensitivity than an EP-based sensor, because the TPD slightly shifts away from the EP~\cite{KottosNature.607.697.2022}, its sensitivity is still constrained by the inherent PF problem.\\

\noindent\textbf {The response enhancement factors}\\
The dynamic response of our CP-based sensor is defined as $R_{\text{CP}, \text{m}} = \partial \Delta\mu_{\text{m}} / \partial B$ with $\text{m} = \text{VP}, \text{AP}, \text{HP}$. Accordingly, the dynamic responses for the VP, AP, and HP measurements are respectively
\begin{subequations}\label{eq:ResponseR}
	\begin{align}
		R_{\text{CP}, \text{VP}}
		&= \frac{4g}{\Delta\mu_\mathrm{VP}}\left[
		(\kappa - gB)
		- \frac{ (\kappa - gB)gB}{\sqrt{\kappa^2 + \left(\kappa - gB\right)^2}}
		+ \sqrt{\kappa^2 + \left(\kappa - gB\right)^2}
		\right]
		, \\
		R_{\text{CP}, \text{AP}}
		&=  \frac{4g}{\Delta\mu_\text{AP}} \left[
		-gB
		+ \frac{g^2B^2}{\sqrt{\kappa^2 + g^2B^2}}
		+ \sqrt{\kappa^2 + g^2B^2}
		\right], \\
		R_{\text{CP}, \text{HP}}
		&=  \frac{4g}{\Delta\mu_\text{HP}} \left[
		-(\kappa + gB)
		+ \frac{ (\kappa + gB)gB}{\sqrt{\kappa^2 + \left(\kappa + gB\right)^2}}
		+ \sqrt{\kappa^2 + \left(\kappa + gB\right)^2}
		\right].
	\end{align}
\end{subequations}
In contrast, the dynamic response of a conventional DP-based sensor is $R_{\text{DP}}=2g$. The response enhancement is defined as $G_{R, \text{m}}=R_{\text{CP}, \text{m}}/R_\text{DP} (\text{m} = \text{VP}, \text{AP}, \text{HP})$. Thus,  we have
\begin{subequations}\label{eq:Renhancement}
	\begin{align}
		G_{R,\text{VP}}
		= &  \frac{2}{\Delta\mu_\mathrm{VP}}
		\left[
		(\kappa-gB)
		- \frac{(\kappa-gB)gB}{\sqrt{\kappa^{2}+(\kappa-gB)^{2}}}
		+ \sqrt{\kappa^{2}+(\kappa-gB)^{2}}
		\right], \\
		G_{R,\text{AP}}
		= &  \frac{2}{\Delta\mu_\text{AP}}
		\left[
		-gB
		+ \frac{g^{2}B^{2}}{\sqrt{\kappa^{2}+g^{2}B^{2}}}
		+ \sqrt{\kappa^{2}+g^{2}B^{2}}
		\right], \\
		G_{R,\text{HP}}
		= &  \frac{2}{\Delta\mu_\text{HP}}
		\left[
		-(\kappa+gB)
		+ \frac{(\kappa+gB)gB}{\sqrt{\kappa^{2}+(\kappa+gB)^{2}}}
		+ \sqrt{\kappa^{2}+(\kappa+gB)^{2}}
		\right].
	\end{align}
\end{subequations}
Theoretically, these response enhancements approaches to  divergent values near the CP, implying the superiority of the CP-based sensors over the DP-based ones in signal enhancement.\\

\noindent\textbf {Fitting of experimental results}\\
In this section, we present the theoretical fittings to the experimental results in the CP sensing without artificial noise as an example. The same method is used for the cases with artificial intensity noise and background noise. Firstly, based on Eq.~\eqref{eq:Dmu}, we fit the experimental SES data in Fig.~\ref{Fig:response}\textbf{a} to extract the coupling strength $g$ and the dissipation rate $\kappa$. The extracted $g$ is $0.90~\text{MHz}/\text{mT}$ for the DP sensor, and $0.89$, $0.89$, and $0.90~\text{MHz}/\text{mT}$ for the VP, AP, and HP measurements of the CP sensor, respectively. The corresponding dissipation rates are $2.34$, $2.50$, and $2.46~\text{MHz}$ for the VP, AP, and HP cases. The close agreement of these values indicates consistency between the experimental observations and theoretical predictions. Therefore, in the subsequent data fitting, we use the extracted coupling strengths and dissipation rates for each measurement case. According to Eqs.~\eqref{eq:Renhancement}, the theoretical curves of response enhancement in  the three measurement scenarios are presented in Fig.~\ref{Fig:response}\textbf{b}. Obviously, the theoretical model is in good agreement with the experimental results.

In our sensor, the sources of noise mainly consist of four parts, namely signal-related noise $\sigma_{\text{signal}}$, shot noise $\sigma_{\text{shot}}$, classical noise $\sigma_{\text{cl}}$, and background noise $\sigma_{0}$, as presented in Eq.~\eqref{eq:Hermitiannoise}. Here, the signal-related noise contribution scales linearly with the magnetic field, namely $\sigma_{\text{signal}}=\gamma_{\text{CP(DP)}} B$, with $\gamma_{\text{CP(DP)}}$ being the fitting parameter. The other three noise contributions are unaffected by the PF and denoted as $\sigma_{\text{other}}$. We assume they are constant for the purpose of fitting. Thus, the expressions for noise amplification and sensitivity are
\begin{equation}\label{eq:noiseamplification}
	G_{\sigma} = \frac{\sigma_{\text{CP}}}{\sigma_{\text{DP}}} = \sqrt{[\sigma_{\text{other}}^{2}+(R_{\text{CP}}\gamma_{\text{CP}} B)^2]/
		[\sigma_{\text{other}}^{2}+(R_{\text{DP}}\gamma_{\text{DP}} B)^2]}\;,
\end{equation}
\begin{equation}\label{eq:sensity}
	S_\text{x}=\frac{\sigma_{\text{x}}}{R_{\text{x}}} \sqrt{\tau} = \sqrt{\sigma_{\text{other}}^{2}/R_{\text{x}}^2+\sigma_{\text{signal}}^2}\sqrt{\tau} \;,
\end{equation}
with $\text{x} = \{\text{CP}, \text{DP}\}$ and $\tau$ being the total measurement time~\cite{natphysics.21.8.2025}. The sensitivity is enhanced by a factor of
\begin{equation}\label{eq:sensityenhance}
	G_S = G_R/G_{\sigma} \;.
\end{equation}
Clearly, when $\sigma_{\text{signal}}$ is negligible, $G_{\sigma} \approx 1$ (see Figs.~\ref{Fig:sensitivity}\textbf{b} and~\ref{Fig:sensitivity}\textbf{c}).
Then, experimental data are fitted using Eqs.~\eqref{eq:noiseamplification}-\eqref{eq:sensityenhance} with the results displayed in Fig.~\ref{Fig:sensitivity}.

\section*{Data Availability}
The data that support the plots within this paper are available via figshare. All other data used in this study are available from the corresponding authors upon reasonable request.

\section*{Acknowledgements}
This work was supported by the Innovation Program for Quantum Science and Technology (Grants No. 2021ZD0301400), the National Natural Science Foundation of China (Grants No. U25A6008, No. 92365107, No. 12305020, No. 12334012, and No. 62575225), the National Key R\&D Program of China (Grants No. 2019YFA0308700 and No. 2019YFA0308704), the Program for Innovative Talents and Teams in Jiangsu (Grant No. JSSCTD202138), China Postdoctoral Science Foundation (Grant No.~2023M731613), and Jiangsu Funding Program for Excellent Postdoctoral Talent (Grant No.~2023ZB708). 

\section*{Author contributions}
K.X. and J.-S.T. conceived the original idea  and the research. K.X. and J.-S.T. co-supervised the project. L.-Q.X. conducted the experiment and data acquisition, and performed data analysis with the help of K.X. and J.-S.T.  W.L. and Y.-Q.L. refined the idea. Y.-P.R., H.Zh., Y.J., H.-D.W. and Y.-Q.L. helped conduction of the experiment. K.X., J.-S.T., and L.-Q.X. built the theoretical model for interpretation of experimental results. K.X. contributed to the presentation of physics and strengthened the theoretical model. All authors contributed to discussion of experimental results and manuscript writing.

\section*{Supplementary Materials}
Supplementary Text (Experimental setup and measurement; Characterization of random technical noise; Sublinear spectral-extremum splitting in the presence of noise; Characterization of orthogonal cavity modes at zero magnetic field; Petermann factor and amplification of noise) \\
Figs. 5 to 10\\
Additional references \cite{proakis2007digital,RevModPhys.82.1155.2010}

\newpage

\centerline{\huge{Supplementary Materials}}

\subsection*{Experimental setup and measurement}
\begin{figure}
	\centering
	\includegraphics[width=0.9\linewidth]{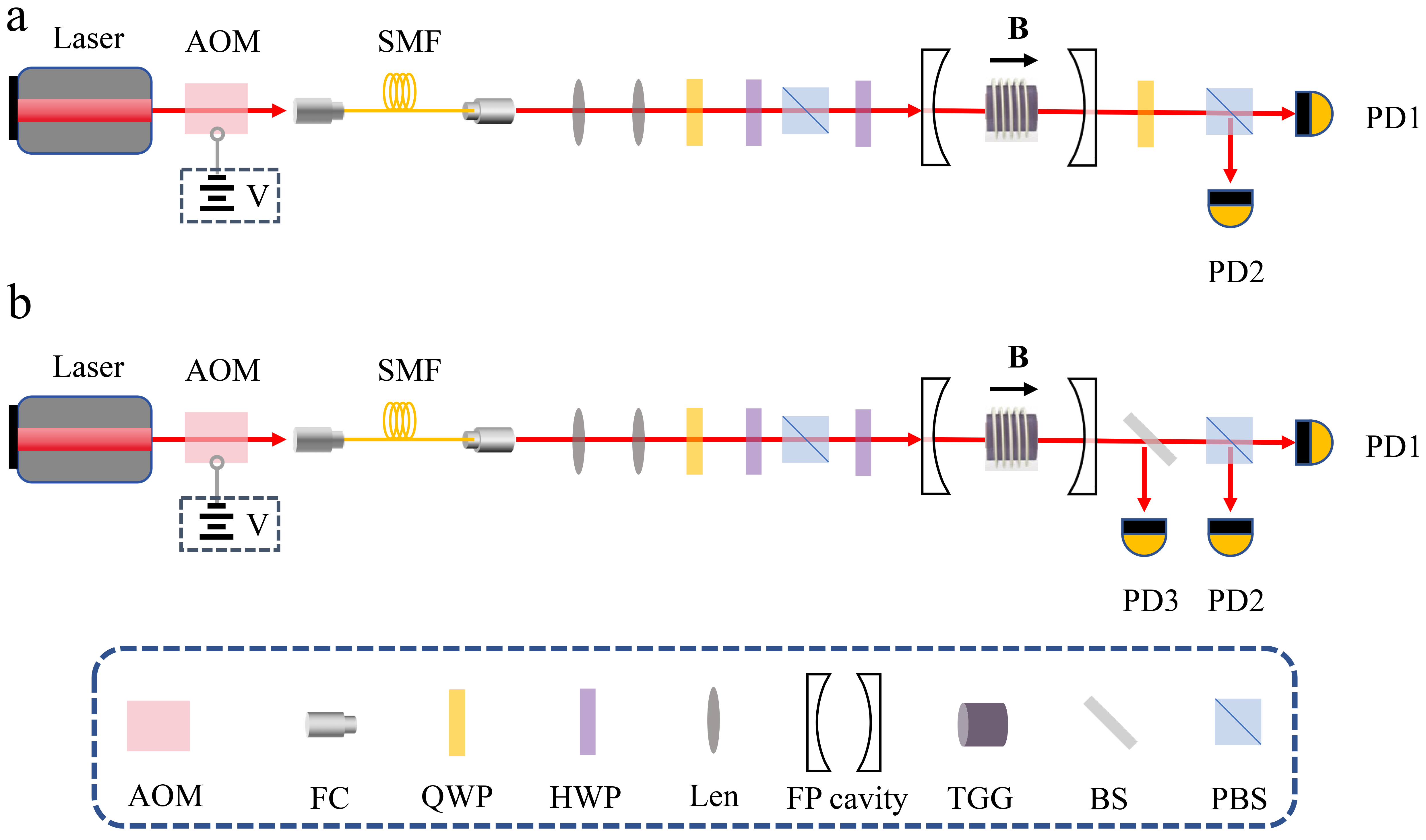} \\
	\caption{Experimental setups of DP- and CP-based sensors. Schematics of experimental setups for (a) the DP-based sensor measuring the eigenmode transmission spectra and (b) the CP-based sensor measuring the transmission spectra in the AP, HP, and VP spaces. FC, fiber coupler; SMF, single mode fiber; QWP, quarter-wave plate; HWP, half-wave plate; PBS, polarization beam splitter; BS, beam splitter; PD, photoelectric detector; TGG, terbium gallium garnet. The acousto-optic modulator (AOM) is used to add artificial intensity noise to the probe laser field.}
	\label{fig:setup1}
\end{figure}

\begin{figure}
	\centering
	\includegraphics[width=0.9\linewidth]{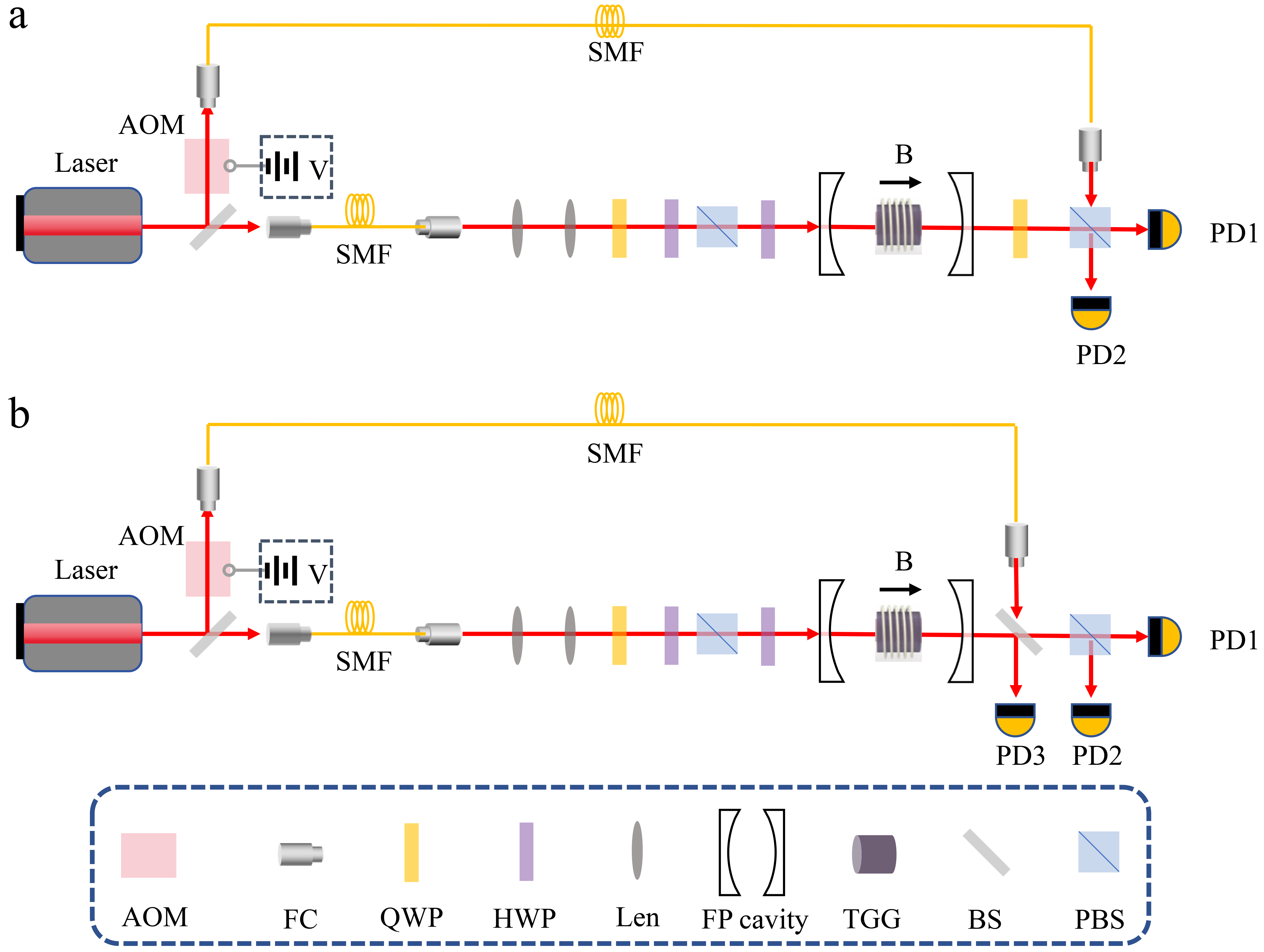} \\
	\caption{Experimental setups for background-noise sensing comparison. Schematic of the experimental setups for (a) the DP-based sensor and (b) the CP-based sensor used to examine the effect of background noise on the sensitivity enhancement. Component representations are the same as in Fig.~\ref{fig:setup1}.}
	\label{fig:setup2}
\end{figure}

The main component of our experiment is a high-quality magneto-optical (MO) Fabry-P\'{e}rot (FP) cavity, as shown in Figs.~\ref{fig:setup1} and \ref{fig:setup2}. A \SI{20}{\milli\metre}-long ($111$)-cut cylindrical terbium gallium garnet (TGG) crystal is placed at the cavity center. To precisely control the magnetic field $B$ and the perturbation, an electronic coil is wound around the TGG crystal and connected to a precision current source (ITECH, IT6302). The probe beam is generated by a tunable external cavity diode laser (Toptica DLC pro), and its intensity is modulated by an acousto-optic modulator (AOM) driven by an applied voltage $V$. A random fluctuation $\delta V$ is on top of a bias voltage $V_0$ to add noise to the system. The bias is fixed at $V_0 = \SI{5.09}{\volt}$. The laser field is then coupled into a \SI{6}{\meter} long single-mode fiber (SMF) for beam shaping and subsequently collimated through a second fiber coupler (FC). The output beam is adjusted by two convex lenses to yield a waist diameter of approximately \SI{225}{\micro\metre} at the cavity center. After passing through a quarter-wave plate (QWP), a half-wave plate (HWP), a polarization beam splitter (PBS), and a second HWP, the beam becomes linearly polarized. Its polarization angle can be adjustable (set to \SI{45}{\degree} in this experiment). The transmitted fields are detected with photodetectors (PDs) and recorded on an oscilloscope for analysis. By sweeping the frequency of the probe laser, we obtain the transmission spectra.

In Fig.~\ref{fig:setup1}, we examine the influence of the laser intensity noise on the sensitivity enhancement. The artificial intensity noise is introduced into the system by applying a random electric voltage signal $\delta V$ to the AOM.
In Fig.~\ref{fig:setup1}a for the DP-based sensor, we measure the eigenmode transmission spectra (Fig.~\ref{fig:setup1}a). A QWP and a PBS together separate the left-circularly polarized (LCP) and the right-circularly polarized (RCP) eigenmodes into two parts, which are subsequently detected by PDs.
In Fig.~\ref{fig:setup1}b, the output field in the CP-based sensor is first separated by a beam splitter (BS) into two paths. One is directly measured as the all-polarization (AP) space by the PD3. The other is further separated by a PBS into the horizontally polarized (HP) and vertically polarized (VP) components, measured by the PD1 and PD2, respectively. In these three measurements, interference between the RCP and LCP components takes place.

Figure~\ref{fig:setup2} schematically illustrates the experimental setups for examining the influence of the background noise on the sensitivity enhancement. A small fraction of the probe laser field is first separated by a BS, then amplitude-modulated with random noise by an AOM, and finally added to the output field to simulate a strong background noise, as in ~\cite{PengXuePRL.133.180801.2024}.

In comparison of noise and sensitivity in the case without artificial noise, we set $\delta V =0$ in Fig.~\ref{fig:setup1}. In practice, the AOM introduce slight noise to the laser field even when $\delta V = 0$ because of the thermal-induced and dark electric noise, as detailed in Sec.~\ref{sec:noise} below. Figure~\ref{fig:noisewithoutAOM} presents the noise amplification measured in the three detection schemes without the AOM inserted after the probe laser. In the absence of additional noise introduced by the AOM, the contribution from signal-related noise becomes stronger. Therefore, for the AP and HP measurements, the noise amplification rises to about $1.6$ near their CP, higher than that when the AOM is inserted (see Fig.~4a in the main text), as indicated by the blue and green data points in Fig.~\ref{fig:noisewithoutAOM}.

\begin{figure}
	\centering
	\includegraphics[width=0.6\linewidth]{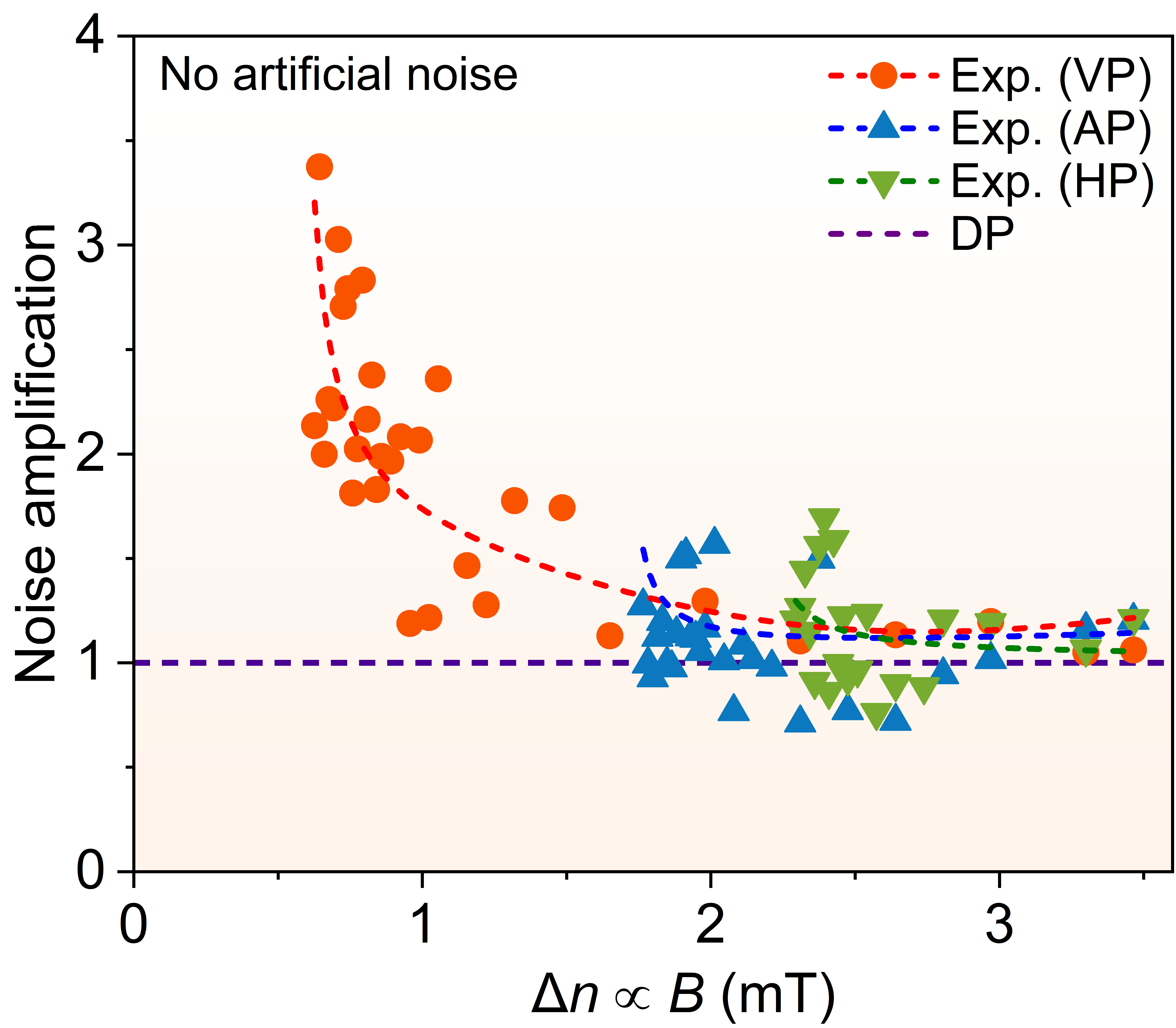} \\
	\caption{Noise amplification of the CP-based sensor relative to the conventional DP-based sensor. The AOM after the probe laser is removed. Orange, blue, and green data points correspond to the VP, AP, and HP measurements, respectively. Experimental data are well fitted (see Methods for details).}
	\label{fig:noisewithoutAOM}
\end{figure}

\subsection*{Characterization of random technical noise}\label{sec:noise}
\begin{figure}
	\centering
	\includegraphics[width=1.0\linewidth]{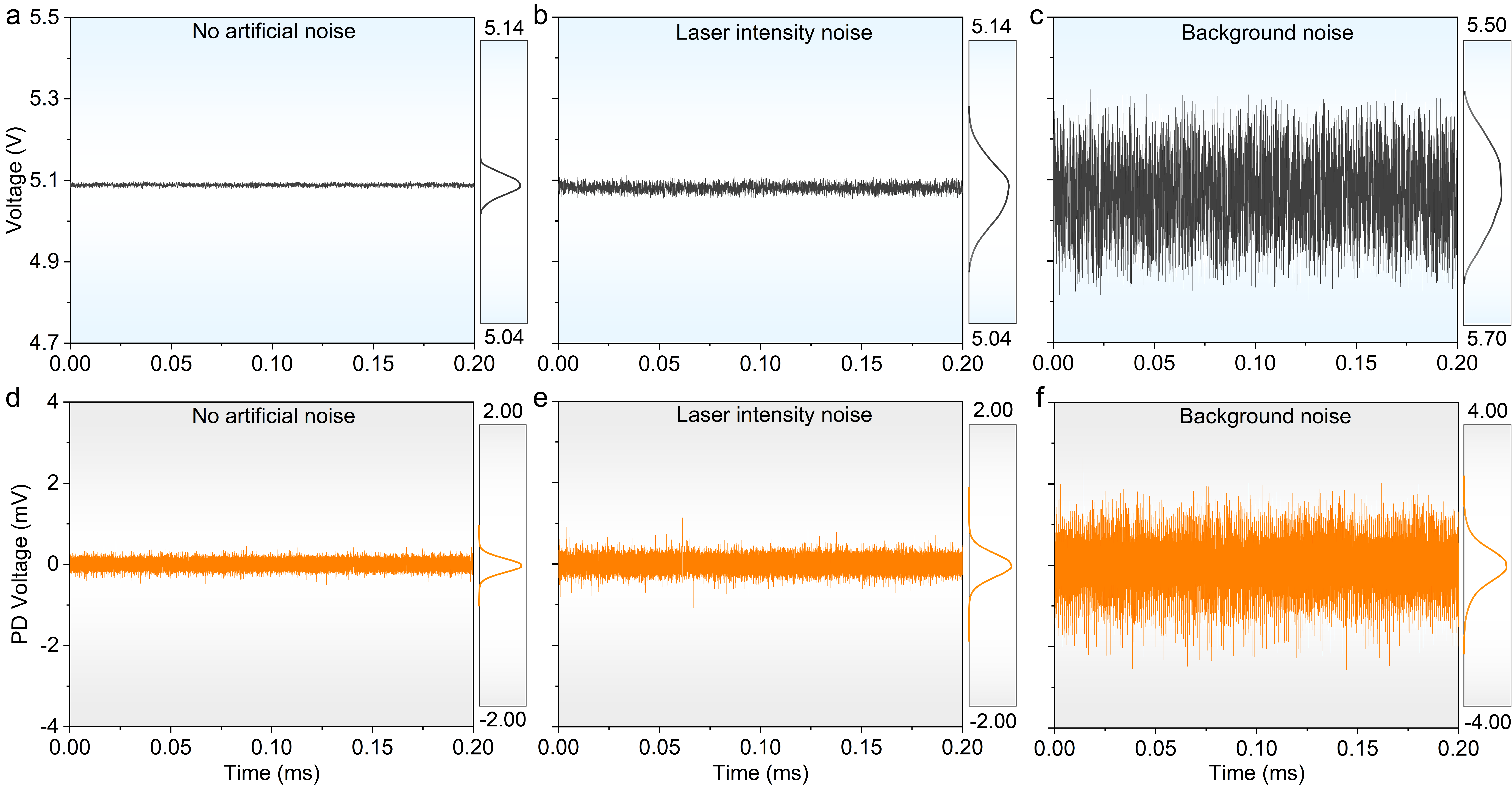} \\
	\caption{Characterization of noise injection. a-c, Driving voltages applied to the AOM for no artificial noise (a), laser intensity noise (b), and background noise (c). d-f, Corresponding optical power signals recorded by PDs after substracting the mean values. The noise distributions are presented on the right panels of the noises.}
	\label{fig:noisecharact}
\end{figure}

In our experiments, random technical noise is artificially introduced via the AOMs. The AOMs are controlled by external electric voltage $V$. The light is diffracted with an efficiency approximately proportional to the applied drive voltage. To ensure reliable long-term operation, the maximum permissible drive voltage applied on our AOMs is up to $V_0 = \SI{5.09}{\volt}$. This value is set as the initial operating point. All noise injection experiments are carried out around this bias. At this operating point, the intrinsic stability of the electronics is first characterized. In the absence of any artificial noise, the control voltage of the AOM still exhibits minor fluctuations, as shown in Fig.~\ref{fig:noisecharact}a. The right panel depicts the random noise distribution with a variance of $\delta V^2 = \SI{1e-5}{\volt\squared}$, which corresponds to the background jitter of the driving source due to thermal and dark electric noise. This small variance indicates the electronic noise floor of the AOM.

To emulate laser intensity noise, a uniformly distributed random modulation is superimposed onto the AOM drive voltage. The additional fluctuations increase the variance to $\delta V^2 = \SI{9e-5}{\volt\squared}$, as shown in Fig.~\ref{fig:noisecharact}b. In this case, the transmitted laser field is subject to white-noise amplitude modulation, thereby providing a controlled setup for testing the PF amplification of these fluctuations.

To investigate the effect of the background noise, an AOM is used to introduce a random noisy optical field before the PDs, as shown in Fig.~\ref{fig:setup2}b. This noisy optical field is controlled by a voltage signal with a larger variance of $\delta V^2 = \SI{1e-2}{\volt\squared}$, as shown in Fig.~\ref{fig:noisecharact}c. It is worth noting that although we apply a random modulation with a uniform distribution to the voltage source, the voltage signal deviates slightly from a perfect uniform distribution due to the presence of intrinsic electronic noise, as shown in the right panel of Fig.~\ref{fig:noisecharact}b. As the modulation amplitude increases, however, the voltage signal gradually approaches a uniform distribution (see the right panel in Fig.~\ref{fig:noisecharact}c).

The optical response is monitored in all three cases with PDs. Since the mean laser power is not strictly constant, we apply a demeaning procedure by subtracting the average detector output, thereby focusing on zero-mean fluctuations~\cite{proakis2007digital}. The processed signals with removing the mean values are displayed in Figs.~\ref{fig:noisecharact}d-f. In the cases of no artificial noise and laser intensity noise, the optical field is detected before entering the optical cavity. At this stage, the modulated light has been collimated and focused through various optical components. The corresponding intensity noise variances measured by the PD are $\SI{0.0089}{\milli\volt\squared}$ and $\SI{0.024}{\milli\volt\squared}$, respectively. In the case of background noise, the optical field is detected at the final detection stage. The intensity noise variance increases significantly to $\SI{0.29}{\milli\volt\squared}$. Therefore, these schemes of adding artificial noise establish a reliable experimental approach for examining the PF-induced noise amplification for both cases with laser intensity noise and background noise.

\subsection*{Sublinear spectral-extremum splitting in the presence of noise}
\begin{figure}
	\centering
	\includegraphics[width=1.0\linewidth]{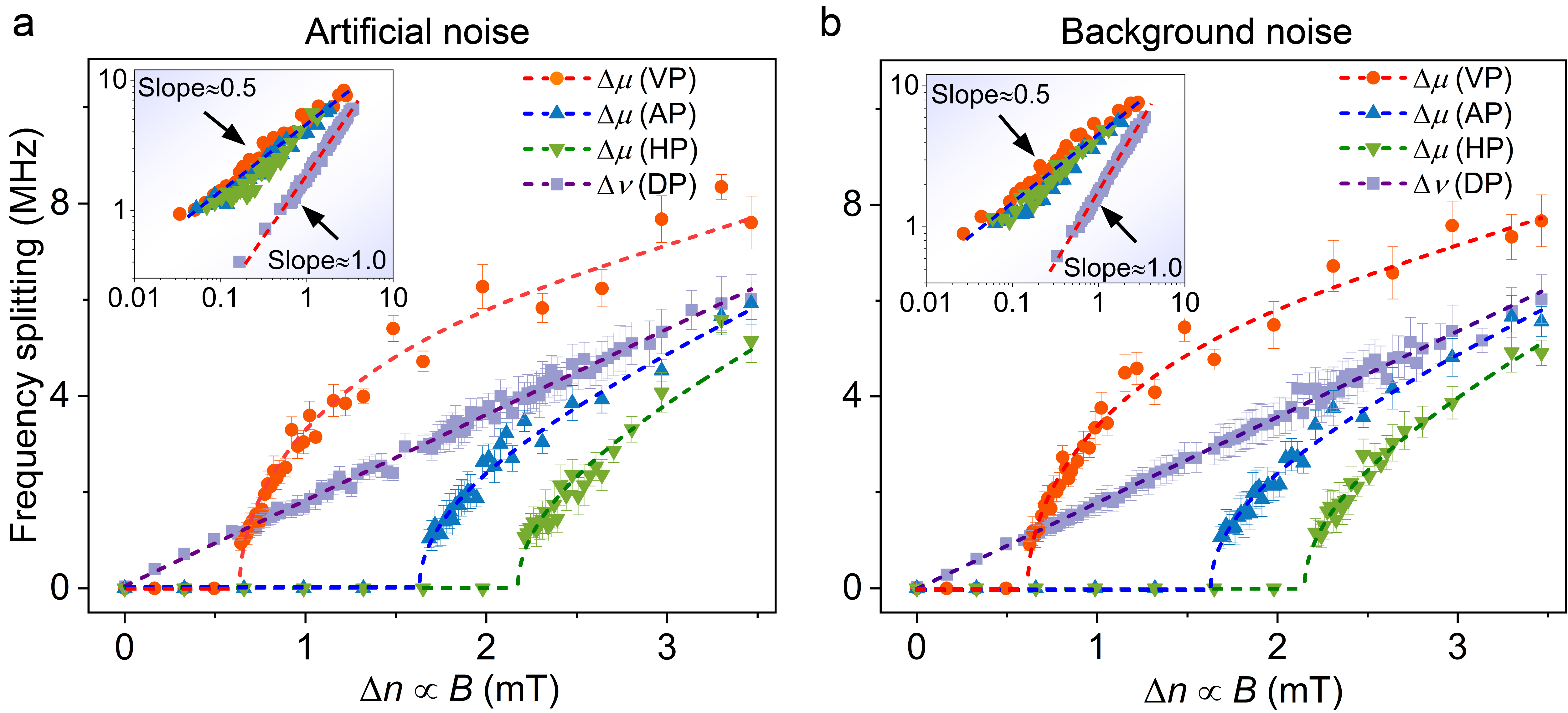} \\
	\caption{Eigenfrequency and spectral-extremum splitting under noise. Eigenfrequency splitting $\Delta \nu$ and spectral-extremum splitting $\Delta \mu$ as a function of the external magnetic field in the cases of applying artificial intensity noise (a) and background noise (b).}
	\label{fig:FSnoise}
\end{figure}

Figure~\ref{fig:FSnoise} compares the eigenfrequency splitting in the DP-based sensor with the spectral-extremum splitting (SES) observed in the CP-based sensor under two noise conditions: intensity noise (Fig.~\ref{fig:FSnoise}a) and background noise (Fig.~\ref{fig:FSnoise}b). The DP sensor exhibits a linear eigenvalue splitting with the perturbation $\Delta n$, proportional to the external magnetic field $B$. In contrast, the CP-based sensor consistently maintains a square-root of the SESs on the perturbation in the VP, AP, and HP measurements. Clearly, the CP-based sensor preserves square-root spectral-extremum splitting even in the presence of noise, in sharp contrast to the linear eigenvalue splitting observed in the DP-based sensor.

To quantitatively determine the frequency splitting, we fit experimental data using the spectral-splitting model given in Eq.~(2). In our experiment, we have $\kappa_\text{H} = \kappa_\text{V} = \kappa$. With fitting, we extract the coupling coefficient $g$ in Eq.~(1) and the decay rate $\kappa$. Under the artificial intensity noise, we obtain the extracted coupling coefficients $g = \{\SI{0.882}{}, \SI{0.874}{}, \SI{0.863}{}\}\si{\mega\hertz\per\milli\tesla}$ and the decay rates of $\kappa = \{\SI{2.403}{}, \SI{2.466}{}, \SI{2.433}{}\}\si{\mega\hertz}$ in the VP, AP, and HP measurements. Under the background noise, the coupling coefficients are extracted to be $g= \{\SI{0.899}{}, \SI{0.874}{}, \SI{0.883}{}\}\si{\mega\hertz\per\milli\tesla}$, while the decay rates are $\kappa =\{\SI{2.369}{}, \SI{2.468}{}, \SI{2.464}{}\}\si{\mega\hertz}$, respectively. These very close coupling coefficients and the decay rates validate our experimental observations.

\subsection*{Characterization of orthogonal cavity modes at zero magnetic field}
\begin{figure}
	\centering
	\includegraphics[width=0.6\linewidth]{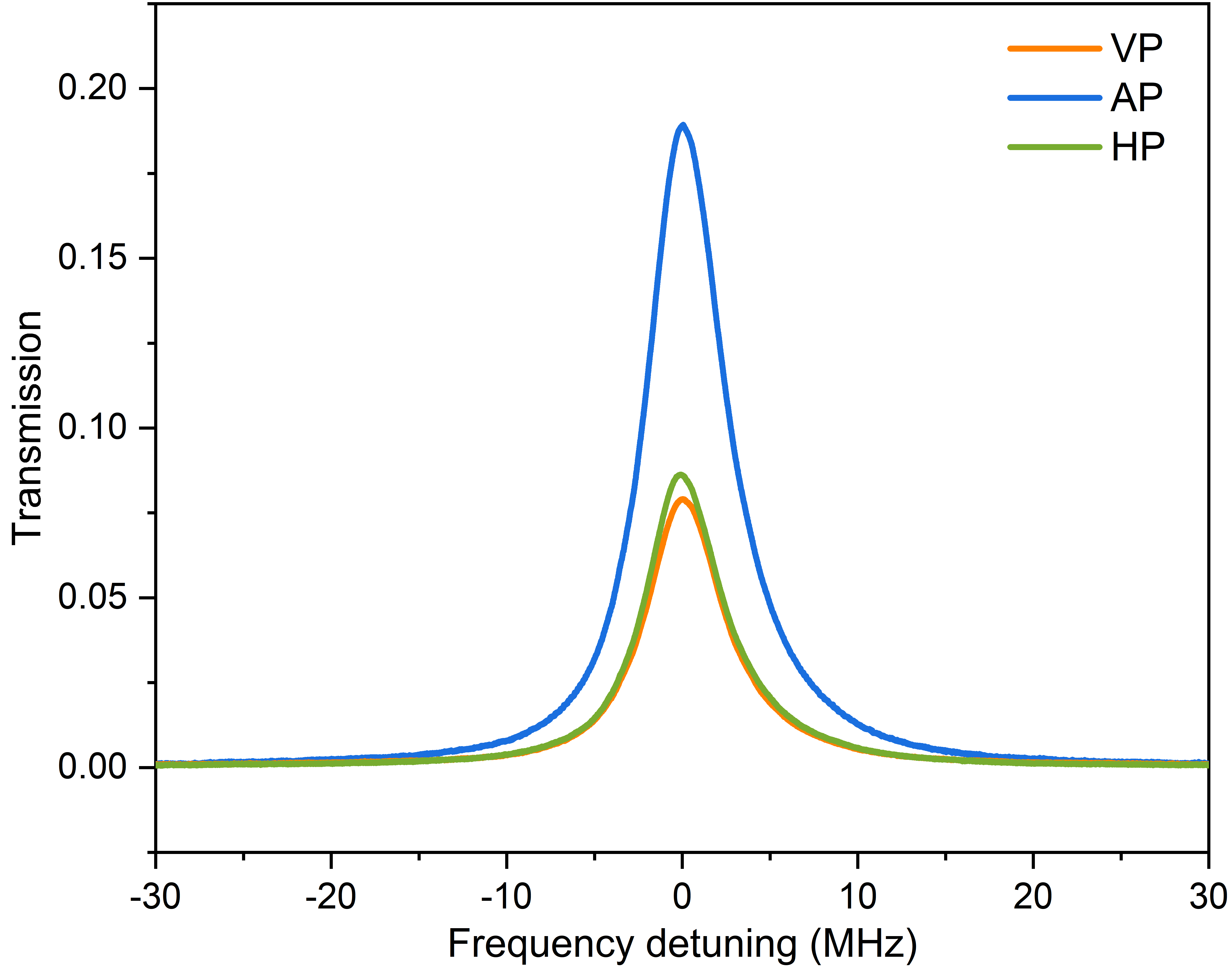} \\
	\caption{Measured transmission spectra at zero magnetic field. Transmission spectra of the HP eigenmode, the VP eigenmode, and the AP mode measured at $B=0$.}
	\label{fig:HVPtransmission}
\end{figure}
In order to clarify the initial conditions of the system, we characterize the HP and VP cavity modes in the absence of external magnetic field ($B=0$). At zero field, the cavity eigenmodes are HP and VP. These two modes are expected to be degenerate in both resonance frequency and linewidth.

The transmission spectra simultaneously measured in the HP, the VP, and the AP spaces are shown in Fig.~\ref{fig:HVPtransmission}. The frequency difference between the HP and VP modes is found to be \SI{0.0069}{\mega\hertz}, which is negligibly small compared to the linewidth. The linewidths are measured to be $\kappa_\text{H}=\SI{5.18}{\mega\hertz}$ for the HP eigenmode and $\kappa_{\text{V}}=\SI{5.24}{\mega\hertz}$ for the VP, respectively. The small difference of \SI{0.06}{\mega\hertz} arises from unavoidable experimental imperfections but can be safely neglected.
These observations confirm that, at $B=0$, the HP and VP eigenmodes are well degenerate in both the resonance frequency and the decay rate, providing a well-defined starting point for subsequent measurement under a finite magnetic field.

\subsection*{Petermann factor and amplification of noise}
Here we provide a detailed derivation of the Petermann Factor (PF) for a general two-mode FP resonator and show explicitly how both quantum and technical drive noise are amplified. Without loss of generality, we assume that the system under consideration supports the HP and VP modes $\hat{a}_{\text{H}}$ and $\hat{a}_{\text{V}}$. These two modes are coupled with a strength $\eta$. For simplicity, we assume that their resonance frequencies are degenerate, equal to $\nu_0$. They decay at rates $\kappa_\text{H}$ and $\kappa_\text{V}$, respectively. The negative (positive) decay rate implies gain (loss). In the frame rotating at the probe frequency $\nu$, the effective Hamiltonian of the coupled system takes the form
\begin{equation}
	\label{seq:Hsys}
	H_c = -\left(\Delta-i\kappa_{\text{H}}\right)\hat{a}_{\text{H}}^{\dagger}\hat{a}_{\text{H}}
	-\left(\Delta-i\kappa_{\text{V}}\right)\hat{a}_{\text{V}}^{\dagger}\hat{a}_{\text{V}}
	+i\eta\left(\hat{a}_{\text{H}}^{\dagger}\hat{a}_{\text{V}}-\hat{a}_{\text{V}}^{\dagger}\hat{a}_{\text{H}}\right),
\end{equation}
where $\Delta=\nu-\nu_0$ is the detuning of the probe field from the bare cavity mode. Removing the trivial trace part yields the reduced non-Hermitian matrix~\cite{VahalaNC.11.1610.2020}
\begin{equation}
	\label{seq:Hsystr}
	H_0 = \begin{bmatrix}
		i \Delta\kappa & i \eta  \\
		- i \eta &  -i \Delta\kappa
	\end{bmatrix}\;,
\end{equation}
with $\Delta\kappa=(\kappa_{\text{V}}-\kappa_{\text{H}})/2$. In the Hermitian case where $\kappa_{\text{H}} = \kappa_{\text{V}}$, this dissipation difference vanishes, i.e., $\Delta\kappa = 0$.

The eigenvalue problem is defined by
\begin{equation}\label{seq:eigenvalue}
	H_0\lvert\psi^R\rangle= \lambda \lvert\psi^R\rangle, \quad \langle\psi^L\rvert H_0= \lambda \langle\psi^L\rvert,
\end{equation}
where $\lvert\psi^R\rangle$ and $\langle\psi^L\rvert$ are the right and left eigenvectors, respectively, and $\lambda$ is the associated complex eigenvalue. Explicitly, for this two-mode system we obtain
\begin{equation}\label{seq:eigenvalueexp}
	\lambda_{\pm}=\pm\sqrt{\eta^2-\Delta\kappa^2}.
\end{equation}
The real part of $\lambda$ corresponds to the mode frequency shift, while the imaginary part describes effective the gain or loss of the eigenmode.

Because the system is non-Hermitian, left and right eigenvectors are not orthogonal under the usual inner product. The non-orthogonality is quantified by the Petermann factor $\text{PF}=\langle\psi^L\lvert\psi^L\rangle\langle\psi^R\rvert\psi^R\rangle$, which for two dimensional matrices gives the closed expression~\cite{VahalaNC.11.1610.2020,KottosNature.607.697.2022}
\begin{equation}\label{seq:PFformula}
	\text{PF}=\frac{1}{2}\left(1+\frac{\text{Tr}\left(H_{0}^{\dagger}H_{0}\right)}{\lvert\text{Tr}\left(H_{0}^{2}\right)\rvert}\right)
	=\frac{1}{2}\left(1+\frac{\eta^2+\Delta\kappa^2}{\lvert\eta^2-\Delta\kappa^2\rvert}\right).
\end{equation}
Clearly, we have $\text{PF}=1$ for Hermitian systems. For non-Hermitian systems at the exceptional point (EP) where $\eta=\Delta\kappa$, the two eigenmodes coalesce and the PF diverges, reflecting maximal non-orthogonality.

To study noise amplification, we write the Langevin equations for the intracavity mode amplitudes. Here, we adopt the vector notation
\begin{equation}\label{seq:vectornotaion}
	\lvert a(t)\rangle=\left[a_{\text{H}}, a_{\text{V}}\right]^T, \quad
	\lvert F(t) \rangle=\left[F_{\text{H}}(t), F_{\text{V}}(t)\right]^T,
\end{equation}
so that the dynamical equation becomes
\begin{equation}\label{seq:Lange}
	\frac{d}{dt}\lvert a(t)\rangle=-iH_0\lvert a(t)\rangle+\sqrt{2\kappa^{\text{ex}}}\alpha\left(t\right)\lvert u\rangle + \lvert F\left(t\right)\rangle.
\end{equation}
The parameter $\kappa^{\text{ex}}$ is the external coupling rate through the input-output mirror, $\lvert u \rangle=\left[\cos\theta, \sin\theta\right]^T$ specifies the polarization of the injected laser field, and $\lvert F(t) \rangle$ represents the vacuum fluctuations associated with gain and loss reservoirs~\cite{nat.commum.9.2018,VahalaNC.11.1610.2020}. This noise satisfies the white-noise correlations $\langle F_{i}(t)F_{j}^{\dagger}(t^\prime)\rangle=\Theta_{i,j}\delta(t-t^\prime)$. The classical drive amplitude is decomposed into a mean value and fluctuations, $\alpha(t)=\alpha_0 + \alpha_{\text{in}}(t)$, where $\alpha_{\text{in}}(t)$ denotes the amplitude noise of the input field. Its power spectral density is $\langle \alpha_{\text{in}}[\Omega]\alpha_{\text{in}}^{*}[\Omega^\prime]\rangle=2\pi\delta\left(\Omega-\Omega^\prime\right)S_{\alpha}(\Omega)$~\cite{RevModPhys.82.1155.2010}.

Projecting onto the left eigenvector $\langle\psi^L\lvert$, the eigenmode amplitude is $c(t)=\langle\psi^L\lvert a(t)\rangle$ and evolves as
\begin{equation}\label{seq:projectleft}
	\dot{c}(t)=-i \lambda c(t) +G_{\text{drv}}(t) + G_{F}(t),
\end{equation}
where $G_{F}(t)=\langle\psi^L\lvert F(t)\rangle$ accounts for quantum noise, and $G_{\text{drv}}(t)=\sqrt{2\kappa^{\text{ex}}}\langle\psi^L\lvert u\rangle \alpha_{\text{in}}(t)$ describes the effective force from input amplitude noise. The corresponding spectra are $S_F(\Omega)=\Theta \langle\psi^L\lvert \psi^L\rangle$ and $S_G(\Omega)=2\kappa^{\text{ex}}\lvert \langle\psi^L\lvert u \rangle\rvert^2S_{\alpha}(\Omega)$. Writing $c(t)=\lvert c \rvert e^{-i\phi}$, the phase dynamics follow 

\begin{equation}
	\dot{\phi}(t) =\frac{i}{2\lvert c\rvert}
	\Bigl[
	\left(G_{\text{drv}}+G_F\right)e^{i\phi}
	-\left(G_{\text{drv}}^{*}+G_{F}^{*}\right)e^{-i\phi}
	\Bigr],
\end{equation}
leading to the noise spectrum
\begin{equation}\label{seq:frenoisespec}
	S_{\varXi}(\Omega)=\frac{1}{2\lvert c\rvert^2}\left[S_F(\Omega)+S_G(\Omega)\right].
\end{equation}
Using the relations $\lvert c\rvert^2=\langle a (t)\lvert a(t)\rangle/\langle\psi^R\lvert \psi^R\rangle$ and $\text{PF}=\langle\psi^L\lvert\psi^L\rangle\langle\psi^R\rvert\psi^R\rangle$, we obtain
\begin{equation}
	S_{\varXi}(\Omega)=\frac{\text{PF}}{2\langle a (t)\lvert a(t)\rangle}\left[\Theta + \frac{2\kappa^{\text{ex}}\lvert\langle\psi^L\lvert u\rangle\rvert^2}{\langle\psi^L\lvert \psi^L\rangle}S_{\alpha}(\Omega)\right].
\end{equation}
This final expression demonstrates that both the fundamental quantum noise $\Theta$ and the technical drive amplitude noise $S_{\alpha}$ are amplified by a factor of PF.


\newpage

\end{document}